\documentclass[reprint,amsmath,amssymb,aps,prb,superscriptaddress]{revtex4-2}

\usepackage{graphicx}% Include figure files
\usepackage{dcolumn}% Align table columns on decimal point
\usepackage{bm}% bold math
\usepackage{siunitx}
\usepackage{multirow}
\usepackage[version=4]{mhchem}
\usepackage[dvipsnames]{xcolor}
 
\usepackage{graphicx}

\newcommand{\greensg}{\mathcal{G}}
\newcommand{\greensf}{\mathcal{F}}
\newcommand{\mop}{\phi}
\def\withfigures{}  % disable bulky figures to speed up recompile

\begin{document}
\preprint{}

\title{Non-local Tunneling Spectroscopy of Inelastic Quasiparticle Relaxation\\ in Superconducting 1-D Wires} %%%%%%%%% Open to Suggestions
% Force line breaks with \\
% \thanks{A footnote to the article title}%

\author{Kevin M. Ryan}
% \altaffiliation[Now At: ]{Department of Physics, Colorado School of Mines, 1523 Illinois Street, Golden, Colorado 80401, USA}
\altaffiliation[Now At: ]{National Institute of Standards and Technology, Boulder, Colorado 80305, USA}
\affiliation{Department of Physics and Astronomy, Northwestern University, 2145 Sheridan Road, Evanston, Illinois 60208, USA}

\author{Detlef Beckmann}
\affiliation{Institute for Quantum Materials and Technologies,
Karlsruhe Institute of Technology (KIT),
Kaiserstraße 12, 76131 Karlsruhe, Germany}

\author{Venkat Chandrasekhar}
\altaffiliation[Contact Author: ]{v-chandrasekhar@northwestern.edu}
\affiliation{Department of Physics and Astronomy, Northwestern University, 2145 Sheridan Road, Evanston, Illinois 60208, USA}%Lines break automatically or can be forced with \\

%  \homepage{http://www.Second.institution.edu/~Charlie.Author}
% \affiliation{
%  Second institution and/or address\\
%  This line break forced% with \\
% }%
% \affiliation{
%  Third institution, the second for Charlie Author
% }%
% \author{Delta Author}
% \affiliation{%
%  Authors' institution and/or address\\
%  This line break forced with \textbackslash\textbackslash
% }%

\date{\today}% It is always \today, today,
             %  but any date may be explicitly specified

\begin{abstract}
Non-local conductance experiments using tunnel junctions can provide valuable spectroscopic information on both the transport and relaxation of quasiparticles in superconductors, as these techniques directly probe the quasiparticle charge and energy imbalance even at mK temperatures. In this work, we employ mesoscopic three terminal \ce{Cu} and \ce{Al} NIS devices to study non-local quasiparticle transport over length-scales on the order of the superconducting coherence length in this regime. Via a dual-bias scheme, which utilizes detector biases both above and below the superconducting gap, we are able to extract the effect of quasiparticle energy imbalance via its impact on the self consistent pair potential by symmetry considerations. We observe non-local conductance features due to pair-breaking which are anti-symmetric with respect to the polarity of the voltage bias, with a sharp onset during single electron tunneling at energies around $3\Delta$. We compare these findings with quasiclassical simulations including inelastic effects to obtain estimates of the energy dependent inelastic scattering time. In addition, we demonstrate kinetic effects due to a large applied supercurrent which can also be captured in this formalism and decomposed with respect to the particle-hole symmetry and supercurrent direction, and discuss further opportunities for the advancement of this method. 
\end{abstract}

%\keywords{Suggested keywords}%Use showkeys class option if keyword
                              %display desired
\maketitle

%\tableofcontents

\section{\label{s:Introduction}Introduction}

When quasiparticles are excited in a superconductor, they contribute to both charge and energy transport over mesoscopic length-scales~\cite{tinkham_introduction_2015}. Such excitations are typically either the result of pair-breaking due to photon or phonon absorption, or via the injection of charge currents from a normal metal or superconductor at a higher chemical potential. By their nature, the properties of these quasiparticles depend significantly on their energy with respect to the superconducting gap, making non-equilibrium quasiparticle transport a sensitive probe of the dynamics and inelastic relaxation of charge carriers deep in the superconducting state. 

Of late, there is a renewed interest in quasiparticle relaxation and recombination due to the emergence of superconducting qubits and other devices which are sensitive to the non-equilibrium quasiparticle density~\cite{arutyunov_relaxation_2018, karzig_quasiparticle_2021, aumentado_quasiparticle_2023}. At the same time, techniques been recently developed to empirically study non-equilibrium quasiparticles at the \SI{}{mK} operating temperatures relevant for these emerging applications. In particular, charge tunneling experiments via normal-metal~\cite{hubler_charge_2010, wolf_spin_2013,ryan_enhanced_2025} or ferromagnetic probes~\cite{kuzmanovic_evidence_2020,beckmann2024} have been essential for establishing charge, energy and spin-imbalance relaxation rates for superconductors in this regime. This is possible via the function of Normal metal-Insulator-Superconductor (NIS) junctions to both inject and detect quasiparticles within the superconductor in the absence of pair-tunneling, as this allows for spectroscopic examination of the underlying non-equilibrium distribution via non-local conductance measurements. 

Here, we report on the quasiparticle imbalance seen via NIS tunneling using a mesoscopic arrangement of tunnel junctions to probe quasiparticle dynamics in a 1D wire of ultra-thin Al. By optimizing the experiment to avoid Joule heating effects within the injector junctions and minimizing the superconducting sample volume, we enable a dual-bias measurement of quasiparticle energy and charge mode imbalance over a length-scale comparable to the dirty-limit superconducting coherence length within the wire. Notably, we are able to discern a sharp drop in the gap during the injection of high-energy quasiparticles which cause pair-breaking during inelastic relaxation.

We analyze this via the particle-hole symmetry of the resulting non-local conductance signal, and explain the effect theoretically via comparison to a quasiclassical model of the non-equilibrium distributions within the wire. By the construction of this device, we are also able to probe kinetic effects under a large applied supercurrent. In those experiments, both pair-breaking and the presence of a phase gradient leads to relaxation and interconversion of the charge and energy mode imbalance of the injected non-equilibrium quasiparticles, which can also distinguished by symmetry with respect to the supercurrent. Finally, we discuss the limitations of our present model with regard to explaining the full anti-symmetric non-local conductance spectra, and make suggestions for future developments of this technique.  

\section{\label{ss:Calculations}Theory}
We begin with a overview of our method for injecting and detecting quasiparticles via NIS tunneling, including a compact description of the underlying quasiclassical theory of non-equilibrium dynamics and superconducting quasiparticle transport. For a deeper discussion of the quasiclassical model for such an experiment, we refer the reader to the extensive literature on this topic (see Refs.~\onlinecite{bergeret2018,heikkila2019,beckmann2024} and references therein). 

To set things up, we consider the simple case of a dirty-limit superconducting wire of cross section $A$ as a function of position $x$. Throughout the paper, we will approximate that the magnitude of the pair potential $|\Delta|$ is constant along the wire for ease of calculations, and similarly we will include a gradient of the order parameter phase $\mop$ only where it is relevant. With these assertions, the spectral properties can be described by the Usadel equation
\begin{equation}
	\Delta \greensg+i\left(\epsilon+i\Gamma\right)\greensf-\zeta\greensf\greensg=0,\label{eqn:usadel}
\end{equation}
where $\greensg$ and $\greensf$ are the normal and anomalous Green's functions, $\epsilon$ is the single-particle energy with respect to the Fermi energy, $\Gamma$ is the lifetime (Dynes) broading parameter, and $\zeta$ is the depairing parameter. We will take all energy-like parameters to be normalized to the pair potential $\Delta_0$ at zero temperature, and the position $x$ along the wire is normalized to the dirty-limit coherence length
\begin{equation}
    \xi=\sqrt{\hbar D_\mathrm{N}/\Delta_0},\label{eq:dirtyxi}
\end{equation}
where $D_\mathrm{N}$ is the normal-state diffusion constant. 

The quasiparticle occupation is described by two distribution functions $f_\mathrm{L}$ (longitudinal) and $f_\mathrm{T}$ (transverse) which are associated with energy and charge respectively~\cite{schmid_linearized_1975}.
At equilibrium, $f_\mathrm{L}=n_0(\epsilon)$ and $f_\mathrm{T}=0$, where $t=k_\mathrm{B}T/\Delta_0$ is the normalized temperature, and 
\begin{equation}
    n_0(\epsilon)=\tanh(\epsilon/2t)
\end{equation}
is related to the usual Fermi-function $f_0$ by $n_0 = 1-2f_0$.

The charge imbalance $Q^*$ is determined by the $f_\mathrm{T}$ mode as
\begin{equation}
    Q^* = 2 e N_0\int N(\epsilon) f_\mathrm{T} \, d\epsilon,
\end{equation}
such that this mode entirely accounts for the imbalance of electron- and hole-like quasiparticles. Here $N_0$ is the density of states at the Fermi energy, and $N(\epsilon)=\mathrm{Re}(\greensg)$ is the normalized density of states of the superconductor. As we will discuss momentarily, a finite $Q^*$ is generated via injection of charge into the superconductor via quasiparticles with $|\epsilon|>\Delta$. This charge is deposited by perturbation of the $f_\mathrm{T}$ mode which lifts the chemical potential for quasiparticles away from the chemical potential of the Cooper pairs by an amount
\begin{equation}
    \mu_{Q^*} = \frac{Q^*}{2e N_0}.
\end{equation}

We can similarly express the supercurrent in this notation as
\begin{equation}
	I_\mathrm{s}=I_0\frac{\partial_x\mop}{2}\int j_\epsilon f_\mathrm{L}d\epsilon,\label{eqn:is}
\end{equation}
where $I_0=\Delta_0/eR_\xi$, $j_\epsilon=\mathrm{Im}(\greensf^2)$ is the spectral supercurrent, $R_\xi=\rho_\mathrm{N} \xi/A$ is the normal-state wire resistance on the length $\xi$, and $\rho_\mathrm{N}$ is the normal-state resistivity. As we will later be interested in the case of finite $\partial_x\mop$, we also note that orbital depairing due to an applied supercurrent\cite{anthore_density_2003,maier_spin-dependent_2023} can be expressed as
\begin{equation}
	\zeta=\frac{1}{2}\left(\partial_x\mop\right)^2.\label{eqn:depairing}
\end{equation}

In this notation, the self-consistent pair potential becomes
\begin{equation}
	|\Delta|=\frac{\lambda}{2}\int_{-\omega_\mathrm{D}}^{\omega_\mathrm{D}}\mathrm{Im}(\greensf )f_\mathrm{L}\,d\epsilon,\label{eqn:sc}
\end{equation}
where $\lambda$ is the coupling constant, and $\omega_\mathrm{D}$ is the Debye frequency. As a result, the non-equilibrium superconducting gap is a function of the longitudinal quasiparticle distribution only.  We note that strictly speaking, Eq.~(\ref{eqn:sc}) should also contain a term with $f_\mathrm{T}$ affecting the order parameter phase $\mop$. Since the tunnel currents $\lesssim 30~\mathrm{nA}$ in the experiment are much smaller than the critical current $I_\mathrm{c}\sim 10~\mathrm{\mu A}$ we neglect this term.

Finally, the propagation of the non-equilibrium modes is described by kinetic equations
\begin{eqnarray}
	\mathcal{D}_\mathrm{L}\partial_x^2f_\mathrm{L} 
	+ j_\epsilon\left(\partial_x\mop\right)\left(\partial_xf_\mathrm{T}\right) 
	& = & \mathcal{I}_\mathrm{L},\label{eqn:kinfL} \\
	\mathcal{D}_\mathrm{T}\partial_x^2f_\mathrm{T}
	+j_\epsilon\left(\partial_x\mop\right)\left(\partial_xf_\mathrm{L}\right) 
	& = & \mathcal{R}_\mathrm{T} f_\mathrm{T},\label{eqn:kinfT}
\end{eqnarray}
where $\mathcal{D}_\mathrm{L,T}$ are spectral diffusion coefficients, $\mathcal{R}_\mathrm{T}$ is the charge relaxation rate, and $\mathcal{I}_\mathrm{L}$ is an inelastic collision integral. In the absence of a supercurrent, the charge and energy modes are decoupled.

We can now consider how to use a pair of NIS junctions to simultaneously perturb the $f_\mathrm{L}$ and $f_\mathrm{T}$ modes via the injection of quasiparticles with $|\epsilon|\geq \Delta$ and measure them at points along the wire. Tunnel injection drives the non-equilibrium distributions at rates of
\begin{equation}
	\mathcal{D}_\mathrm{L,T}\partial_x f_\mathrm{L,T} = \kappa_\mathrm{I}N(\epsilon)\delta f_\mathrm{L,T},
\end{equation}
where $\kappa_\mathrm{I}=G_\mathrm{inj} R_\xi$, $G_{inj}$ is the normal-state tunnel conductance, and $\delta f_\mathrm{L,T}$ is the difference of $f_\mathrm{L,T}$ across the junction. Treating the normal-metals as equilibrium reservoirs, such that the distribution functions for electrons on the normal side of the injector and detector junction at bias $V$ are simply given by
\begin{equation}
    f_\mathrm{L,T} = \frac{1}{2}\left(n_0(\epsilon+eV)\pm n_0(\epsilon-eV)\right),\label{eqn:bias}
\end{equation}
we can obtain that the charge current injected is given by
\begin{equation}
	I_\mathrm{inj} = -\frac{G_\mathrm{inj}}{2e} \int N \delta f_\mathrm{T} \,d\epsilon.\label{eqn:tunnel}
\end{equation}
This includes both the bias-driven local current and the non-equilibrium driven non-local current. In most experiments of this type, $G_\mathrm{inj}$ is chosen to be large enough to effectively modify the quasiparticle distributions, but small enough to avoid inducing a significant supercurrent due to recombined pairs such that $\partial_x\mop\approx 0$. Using a second junction as our detector, we then expect to observe a similar current
\begin{equation}
	I_\mathrm{det} = -\frac{G_\mathrm{det}}{2e} \int N(V_\mathrm{inj},V_\mathrm{det})\left(f^\mathrm{(S)}_\mathrm{T}(V_\mathrm{inj})-f^\mathrm{(N)}_\mathrm{T}(V_\mathrm{det}) \right)d\epsilon,\label{eqn:Inonlocal}
\end{equation}
where we have written the dependence on bias conditions explicitly. Here, $f^\mathrm{(N)}_\mathrm{T}(V_\mathrm{det})$ is the distribution in the detector contact given by Eq.~(\ref{eqn:bias}), and  $f^\mathrm{(S)}_\mathrm{T}(V_\mathrm{inj})$ is the distribution function in the superconductor which can be obtained by solving the kinetic equations. Note that the injection rate for $f_\mathrm{L}$, i.e., the heat current through the junction, does not affect the observed \textit{charge} current directly. It does, however, affect the self-consistent pair potential $\Delta$, and therefore the density of states $N$. Together with a supercurrent, it also drives the coupling term in the kinetic equations, and makes the detector junction a probe of the non-equilibrium $f_\mathrm{L}$ when $V_\mathrm{det}\neq0$. 

This construction allows us to identify a non-local conductance between two NIS junctions, which occurs when the injector junction drives the system sufficiently out of equilibrium that a change in the current through the detector junction occurs. From this, we will define the nonlocal conductance
\begin{equation}
 g_\mathrm{nl}=\frac{dI_\mathrm{det}}{dV_\mathrm{inj}}.
\end{equation}
For clarity, when referring to the \textit{local} conductance of a junction, we will explicitly denote this via the uppercase $G(V)$. By our construction, $g_\mathrm{nl}$ contains contributions which are distinguished by their symmetry with respect to  bias voltages and supercurrent. 

To see why, we must first note that the symmetry of $g_\mathrm{nl}$ with respect to $V_\mathrm{inj}$ is opposite to that of $I_\mathrm{det}$ from Eq.~\ref{eqn:Inonlocal} after differentiation. With this, as the total charge imbalance $\mu_{Q^*}$ depends on the charge of the injected carriers, we see that the $f_\mathrm{L}$ mode yields a term in $g_\mathrm{nl}$ which is \textit{even} with respect to $V_\mathrm{inj}$. Conversely, as the self-consistent $\Delta$ depends on $f_\mathrm{L}$ (heat injection), and is therefore \textit{even} in $V_\mathrm{inj}$, one finds that the resulting non-local conductance via $f_L$ mode imbalance must be \textit{odd} with respect to $V_\mathrm{inj}$. Then, considering the potential for intermixing between modes, one finds that the coupling terms in the kinetic equations from Eqs.~\ref{eqn:kinfL} and \ref{eqn:kinfT} are \textit{odd in supercurrent}, while all other terms depend on the supercurrent only indirectly via the depairing rate (Eq.~\ref{eqn:depairing}) and are therefore \textit{even} in the applied supercurrent, yielding an additional distinct symmetry.

Taken altogether, the bias-driven contribution of charge imbalance to $g_\mathrm{nl}$ is \textit{even in injector bias and in supercurrent}, while that of energy imbalance part is \textit{odd in injector (and detector) bias} as well as \textit{even in supercurrent}. The intermixing of the two modes via the kinetic equations is then \textit{odd in bias and in supercurrent}. With this established, it is natural then to decompose the nonlocal conductance as follows:
\begin{eqnarray}
	g_\mathrm{a} & = & \frac{1}{2}\left(g_\mathrm{nl}(I_\mathrm{s})-g_\mathrm{nl}(-I_\mathrm{s})\right), \\
	g_\mathrm{s} & = & \frac{1}{2}\left(g_\mathrm{nl}(I_\mathrm{s})+g_\mathrm{nl}(-I_\mathrm{s})\right), \\
	g_\mathrm{sa} & = & \frac{1}{2}\left(g_\mathrm{s}(V_\mathrm{inj})-g_\mathrm{s}(-V_\mathrm{inj})\right), \\
	g_\mathrm{ss} & = & \frac{1}{2}\left(g_\mathrm{s}(V_\mathrm{inj})+g_\mathrm{s}(-V_\mathrm{inj})\right).
\end{eqnarray}
These symmetries can then be used to separate these effects in experiment, as we will consider later.

\section{\label{s:Methods}Methods}
\begin{figure}[b!]
    \centering
    \ifdefined\withfigures
    \includegraphics[width=0.95\linewidth]{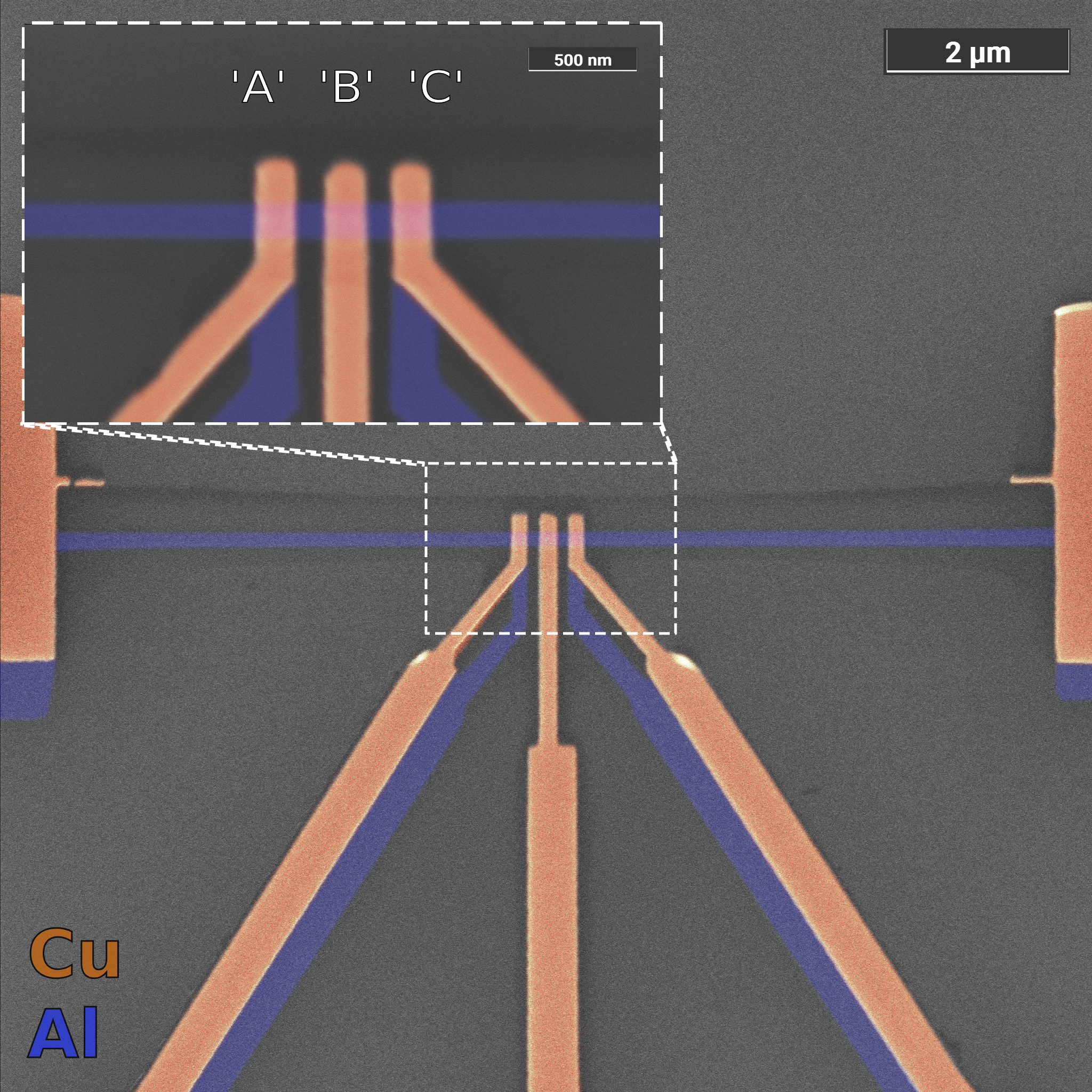}
    \fi
    \caption{False Color SEM Micrograph of one device. The central superconducting wire was isolated via a liftoff step which prevented the formation of a parallel normal metal wire between the left and right reservoirs of the device. Traces of this liftoff process can be seen on either edge where this Cu feature is broken off. In the inset, we label the junctions $A$, $B$, and $C$ as discussed in the text.}
    \label{fig:FalseColor}
\end{figure}
Three terminal NIS junction devices were fabricated via a conventional PMGI/PMMA based liftoff procedure on \SI{1}{\micro\meter} \ce{SiOx} on p-type Si substrates, using a tilting stage electron-beam evaporator with a base pressure of \SI{4E-7}{Torr}. To construct the central superconducting wire of the device, \SI{8}{\nano\meter} of \ce{Al} was deposited from a 99.999\% purity source  at a rate of \SI{0.5}{\AA \per \second}. The Al was then oxidized in flowing \ce{O2} gas at a pressure of \SI{100}{\milli  Torr} for 15 minutes. To form the counter electrodes, \SI{80}{\nano\meter} of 99.999\% purity \ce{Cu} was deposited at a $35^\circ$ angle over the Al wire, which was then followed by a \SI{5}{\nano\meter} Ti layer to passivate the Cu during liftoff and sample handling. By tuning the PMGI undercut, we are able to prevent the formation of a parallel Cu wire across the lateral extent of the device, as this feature is removed during the liftoff process. The junctions (and \ce{Al} wire itself) are nominally \SI{150}{\nano\meter} across and are spaced by \SI{300}{\nano\meter} between centers. Each of the junctions connects to a single measurement lead, while either side of the $l=\SI{11}{\micro\meter}$ long, $w=\SI{150}{\nano\meter}$ wide Al wire are connected to pairs of much wider leads to facilitate four-terminal measurements of the Al and provide strongly superconducting reservoirs. Anecdotally, we find that such arrangements where the overall number of leads is kept small tend to be less sensitive to electromagnetic noise and show less thermal broadening.

\begin{table}[t!]
\resizebox{\columnwidth}{!}{%
\begin{tabular}{cccccccc}
\cline{1-6}
\multicolumn{1}{|c|}{Junction} &
  \multicolumn{1}{c|}{$G_\mathrm{inj}$} &
  \multicolumn{2}{c|}{$\Gamma$} &
  \multicolumn{1}{c|}{$\Delta$} &
  \multicolumn{1}{c|}{$T_e$} &
   &
   \\ \cline{1-6}
\multicolumn{1}{|c|}{A} &
  \multicolumn{1}{c|}{\SI{24.6}{\micro S}} &
  \multicolumn{2}{c|}{$\leq\SI{1E-5}{\micro eV}$} &
  \multicolumn{1}{c|}{\SI{266}{\micro eV}} &
  \multicolumn{1}{c|}{\SI{61}{mK}} &
   &
   \\ \cline{1-6}
\multicolumn{1}{|c|}{B} &
  \multicolumn{1}{c|}{\SI{31.4}{\micro S}} &
  \multicolumn{2}{c|}{$\leq\SI{1E-5}{\micro eV}$} &
  \multicolumn{1}{c|}{\SI{259}{\micro eV}} &
  \multicolumn{1}{c|}{\SI{37}{mK}} &
   &
   \\ \cline{1-6}
\multicolumn{1}{|c|}{C} &
  \multicolumn{1}{c|}{\SI{24.8}{\micro S}} &
  \multicolumn{2}{c|}{$\leq\SI{1E-5}{\micro eV}$} &
  \multicolumn{1}{c|}{\SI{270}{\micro eV}} &
  \multicolumn{1}{c|}{\SI{86}{mK}} &
   &
   \\ \cline{1-6}
\multicolumn{8}{c}{} \\ \cline{1-2}
\multicolumn{2}{|c|}{Al Properties} &
  \multicolumn{6}{c}{} \\ \hline
\multicolumn{1}{|c|}{$R$} &
  \multicolumn{1}{c|}{$l$} &
  \multicolumn{1}{c|}{$w$} &
  \multicolumn{1}{c|}{$l_{mfp}$} &
  \multicolumn{1}{c|}{$D_N$} &
  \multicolumn{1}{c|}{$L_T$} &
  \multicolumn{1}{c|}{$\xi$} &
  \multicolumn{1}{c|}{$J_{c0}$} \\ \hline
\multicolumn{1}{|c|}{$\SI{1.45}{\kilo \ohm}$} &
  \multicolumn{1}{c|}{\SI{11}{\micro \meter}} &
  \multicolumn{1}{c|}{\SI{150}{\nano\meter}} &
  \multicolumn{1}{c|}{\SI{2.49}{\nano\meter}$^\dagger$} &
  \multicolumn{1}{c|}{\SI{16.9}{\centi\meter\squared\per\second}$^\dagger$} &
  \multicolumn{1}{c|}{\SI{803}{\nano\meter}$^\dagger$} &
  \multicolumn{1}{c|}{$\SI{65}{\nano\meter}^\dagger$} &
  \multicolumn{1}{c|}{\SI{1.7}{\mega \ampere\per\centi\meter\squared}$^\dagger$ }\\ \hline
\end{tabular}%
}
\caption{Properties of the NIS junctions and of the central Al wire. Junction values were obtained by fitting to a Dynes-type DOS as described in the main text. In all cases, the Dynes parameter $\Gamma$ is negligible. Properties of the Al wire were determined from the normal state resistance measured above the critical current at \SI{23.5}{mK}. $^\dagger$Due to uncertainty in the Al thickness, the deposited thickness $d=\SI{8}{\nano\meter}$ has been used to estimate a lower bound for these properties.}
\label{tab:Junctions}
\end{table}

After liftoff, a pair of these devices were mounted to a non-magnetic printed circuit board using \ce{Ag} paint and wire-bonded before being loaded onto to an Oxford MX-100 dilution refrigerator for measurement. Continuity of the sample leads was carefully considered to prevent spurious NIS junctions, which we find can be simply avoided by using a single metalization step for both the device and the wirebonding pads. In this way, the Al wirebonds make galvanic contact to both the Al and Cu layers in each lead. Unless otherwise stated, measurements were performed at a temperature of \SI{24}{\milli\kelvin}, as reported by a \ce{RuOx} thermometer on the mixing chamber of the refrigerator. Thermalization of the device was accomplished via strong thermal anchoring of the measurement leads to each stage of the refrigerator, and it was further shielded by a copper radiation shield affixed to the cold plate. Room temperature low-pass $\pi$-filters with a cutoff of \SI{800}{\kilo \hertz} and $\mu$-metal cable shielding were used to further suppress electromagnetic radiation. In the following results, we limit our discussion to a single device which demonstrated the least thermal broadening.% at a base temperature of \SI{20.5}{mK}.

Battery powered home-made summing circuits and ADA4530 based trans-impedance amplifiers (TIAs) were used to perform differential conductance measurements in combination with low noise Lock-in Amplifiers (LIAs). These TIAs were designed to be individually voltage biased and operate at a gain of \SI{1E8}{\volt\per\ampere}. Typical excitation voltages were on the order of \SI{5}{\micro\volt} at frequencies $f\leq \SI{20}{\hertz}$ for all measurements unless otherwise noted, and all pre-amplifiers were located within a vibration isolated $\mu$-metal shielded cabinet. In local conductance measurements, the Al wire running across the sample is grounded on both sides, and a voltage is applied by the TIA to measure the $I-V$ response as well as the local conductance $G_i \equiv dI_i/dV_i$, where the subscript $i$ denotes a specific junction in the three terminal array. For non-local conductance, two TIAs are used to simultaneously bias two NIS junctions with the third junction left disconnected from ground. This allows for simultaneous measurement of the local conductance of the injector, and the non-local conductance of the detector, as well as the dc-currents through both junctions. This particular non-local measurement scheme is closely based on those of Ref.~\onlinecite{hubler_charge_2010, kolenda_nonlocal_2013}, and follow the same sign convention from those works, such that current coming from the detector due to a positive injector voltage yields a positive non-local conductance.

\section{\label{s:Results}Experiment}

\begin{figure}[b!]
    \centering
    \ifdefined\withfigures
    \includegraphics[width=.95\linewidth]{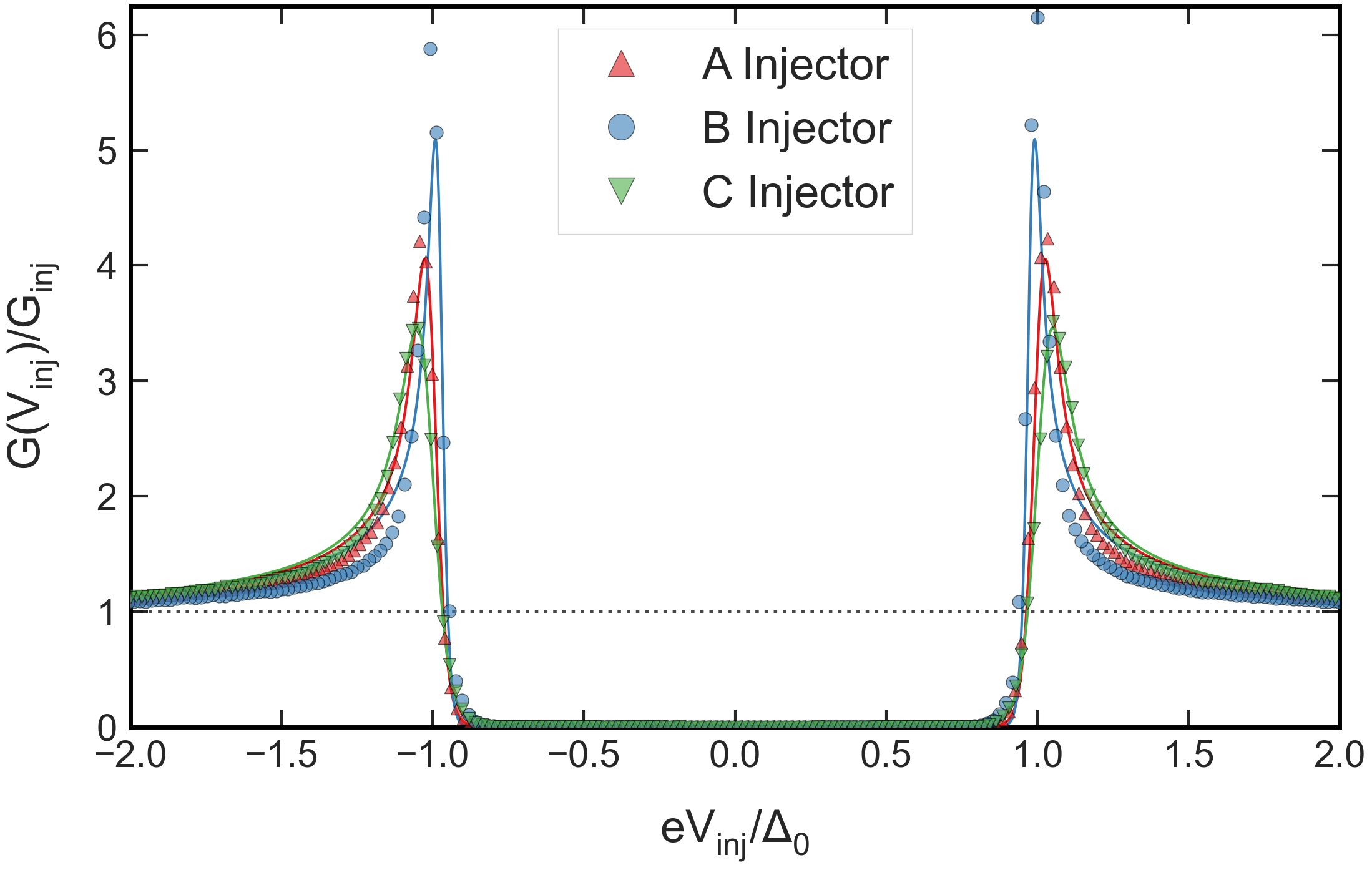}
    \fi
    \caption{Normalized local conductance of the three junctions studied, measured at \SI{20.5}{mK} for all three junctions with the remaining two junctions floating. The normalization values $G_\mathrm{inj}$ and $\Delta_0$ were obtained via fitting as described in the text.  }
    \label{fig:LocalConductance}
\end{figure}

For each NIS junction on the device, the local conductance was measured to ensure a strong tunneling behavior and observe the local density of states along the wire. The conductance spectra could then be fit to a Dynes-type model of the superconducting density of states~(DOS)~\cite{dynes_direct_1978,meservey_spin-polarized_1994}  and extract the effective electronic temperature $T_e$ of the normal metal probes, as reported in Tab. \ref{tab:Junctions}. In all cases, a well gapped spectra was observed without the need for a Dyne's energy $\Gamma$, with only the height of the coherence peak maxima varying between the outer ($A$ and $C$) and inner ($B$) junctions due to thermal broadening.  We posit that this higher than expected electronic temperature on the $A$ and $C$ junctions is related to the construction of the leads: either the outer two electrodes are acting to shield the inner one from stray electromagnetic radiation, or are intrinsically more sensitive due to the un-covered \ce{Al} sections in the vicinity of the junction or due to their slightly lower normal state conductances. As a result of this asymmetric behavior, we have predominantly used junction $B$ as a detector of the non-local quasiparticle conductance in the subsequent findings.

The average value of $\Delta$ among these fits can be taken as the equilibrium value $\Delta_0 = \SI{265\pm5}{\micro eV}$ for the entire wire. From this value, we can estimate the clean-limit coherence length $\xi_0 \approx \hbar v_f/\Delta_0$ to be $\SI{5.0}{\micro\meter}$. By additional measurements of the wire resistance in the normal state via application of a large current across the \ce{Al}, we can also determine the normal state diffusion constant $D_N$ and compute the Thouless length $L_T = \sqrt{ \hbar D_N /k_B T}$ at \SI{24}{mK} to be \SI{733}{\nano\meter}, as well as the dirty-limit value $\xi = \SI{65}{\nano\meter}$ by Eq.~\ref{eq:dirtyxi}. Consequently the \ce{Al} wires are expected to be reasonably 1-D and the junction's spacing small enough to observe non-equilibrium quasiparticle distributions just after injection. These values are also reported in Tab.~\ref{tab:Junctions}, where we additionally report the electronic mean free path $l_{mfp}$ and an estimated upper bound for the equilibrium critical current density $J_{c0}$.

\begin{figure}[t!]
    \centering
    \ifdefined\withfigures
    \includegraphics[width=0.95\linewidth]{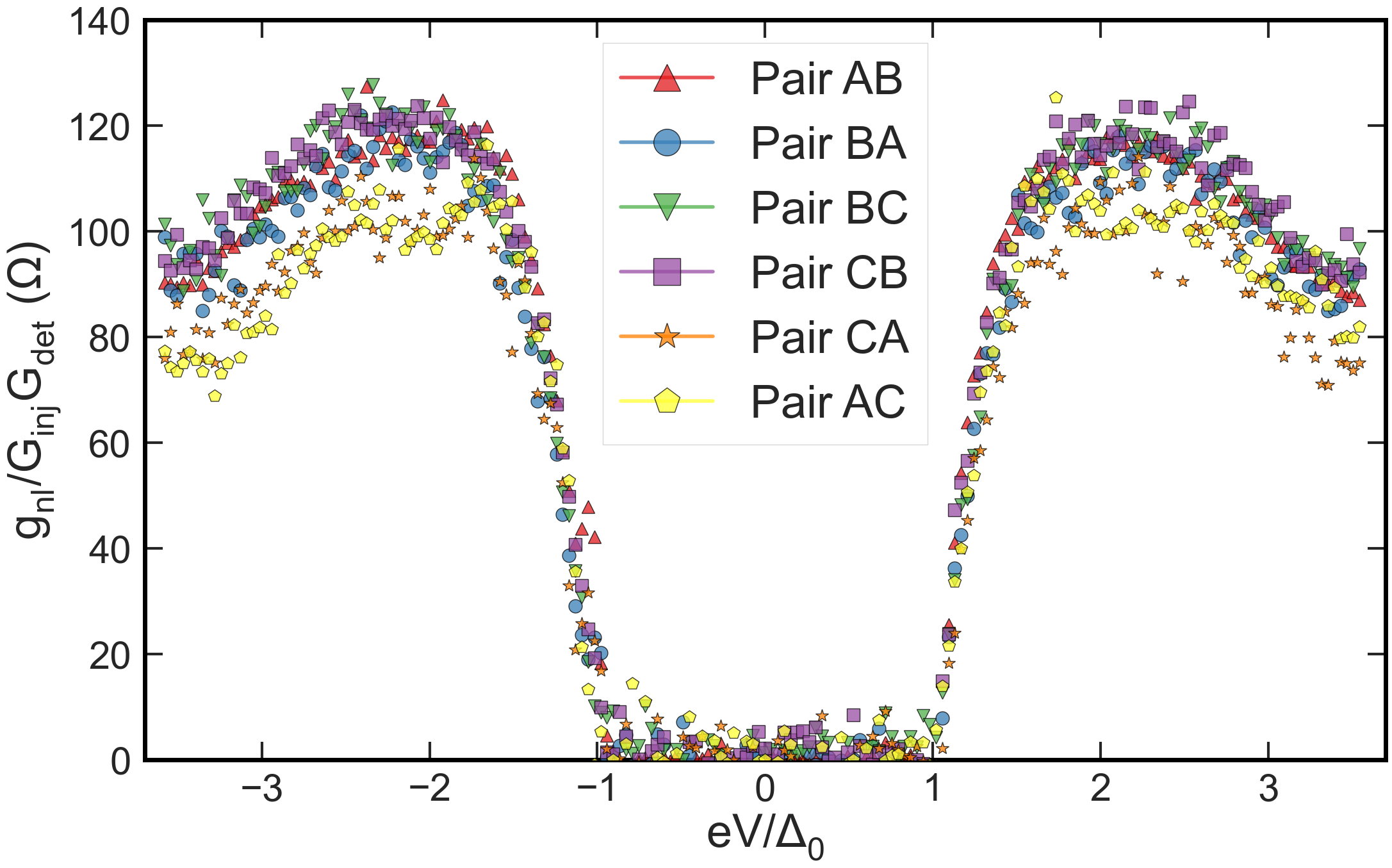}
    \fi
    \caption{Normalized non-local conductance due to charge-imbalance for all combinations of detector and injector. As described in the text, the spacing between the $A$ and $C$ junctions is approximately twice that of the $A/B$ or $B/C$ injector/detector pairs, which can be seen in the reduction of overall charge imbalance signal.}
    \label{fig:ChargeImalanceSignal}
\end{figure}

\begin{figure*}[t!]
    \centering
        \ifdefined\withfigures
    \includegraphics[width=0.48\linewidth]{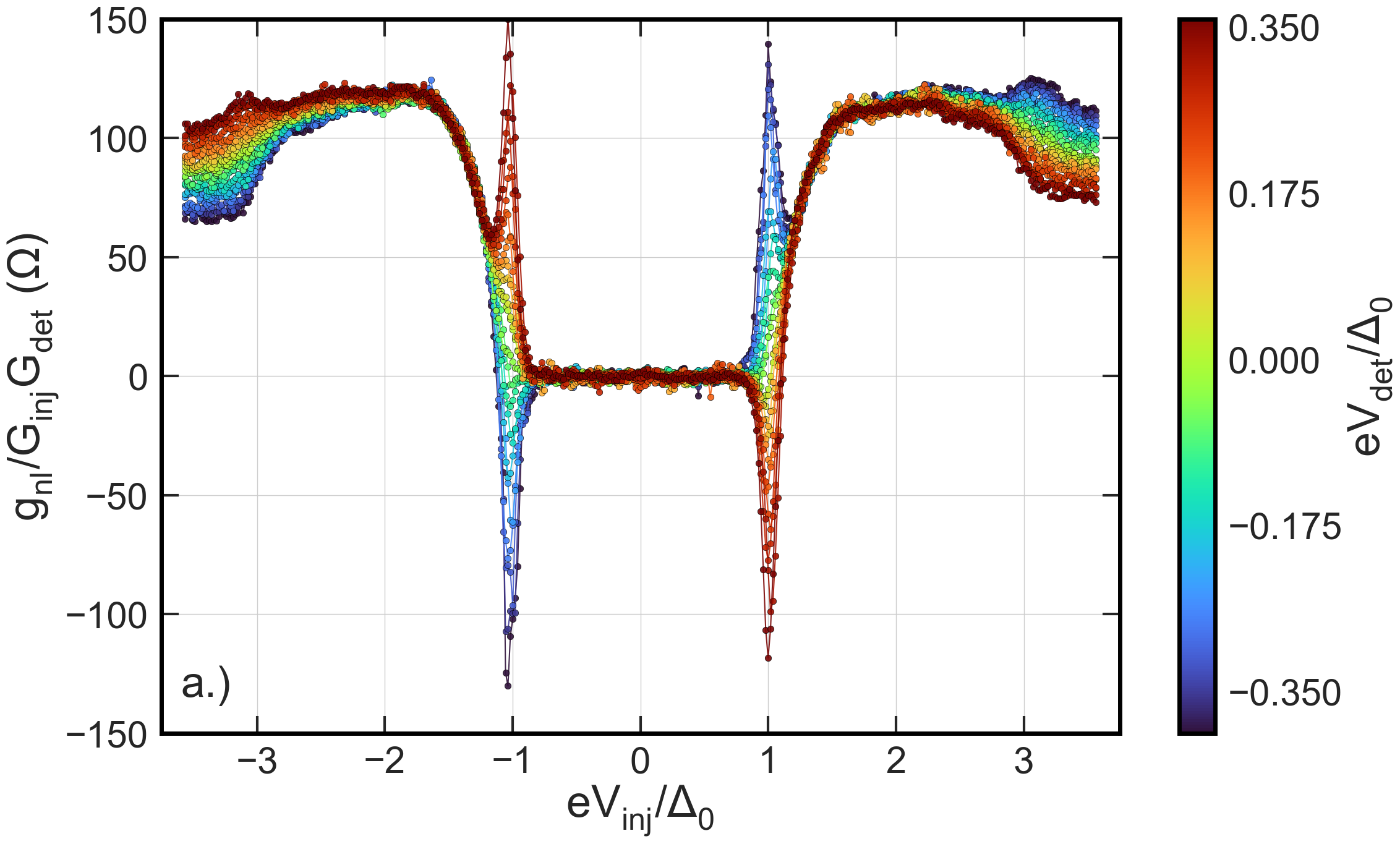}\hfill
    \includegraphics[width=0.48\linewidth]{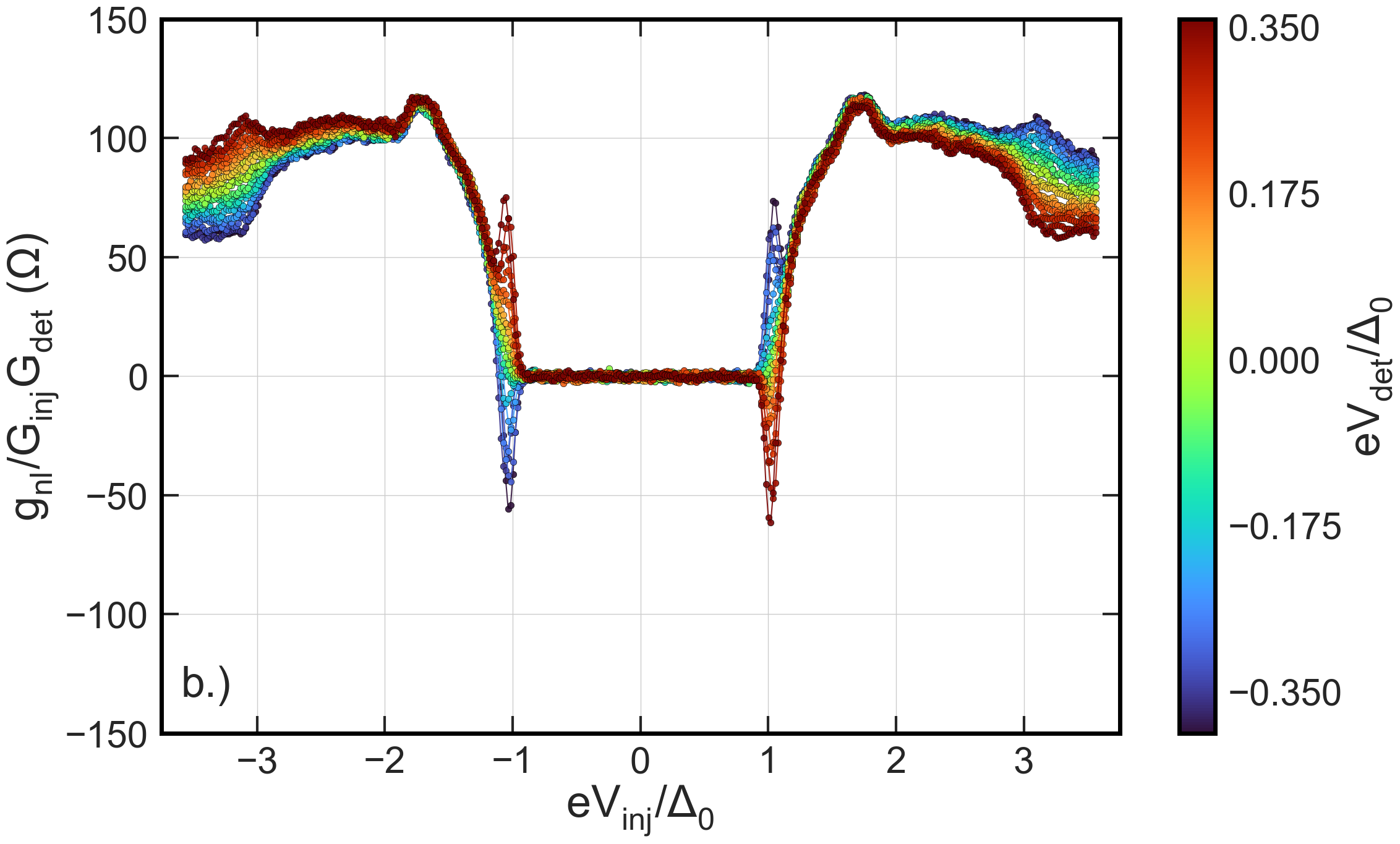}\\
    \vspace{1em}
    \includegraphics[width=0.48\linewidth]{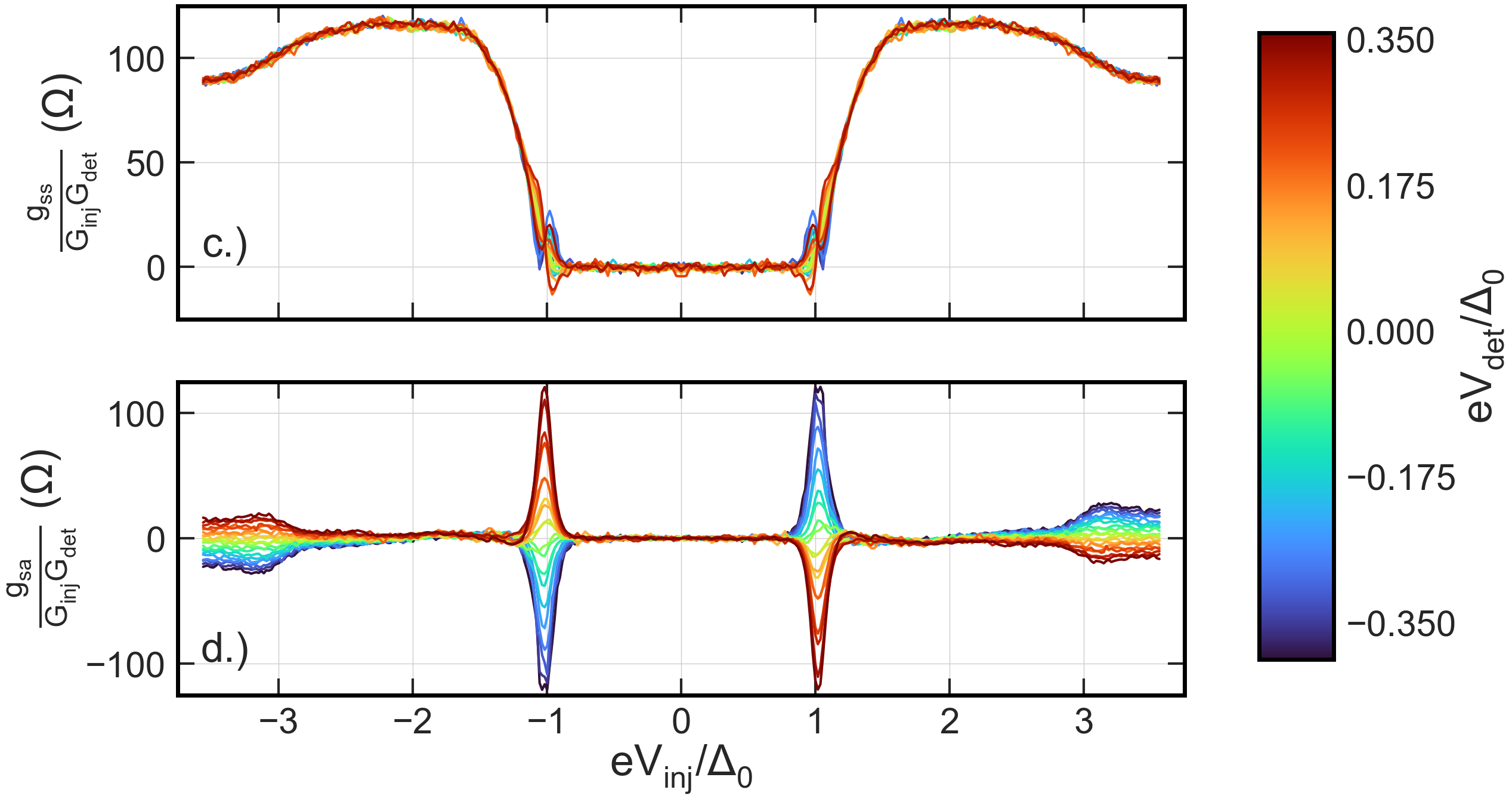}\hfill
    \includegraphics[width=0.48\linewidth]{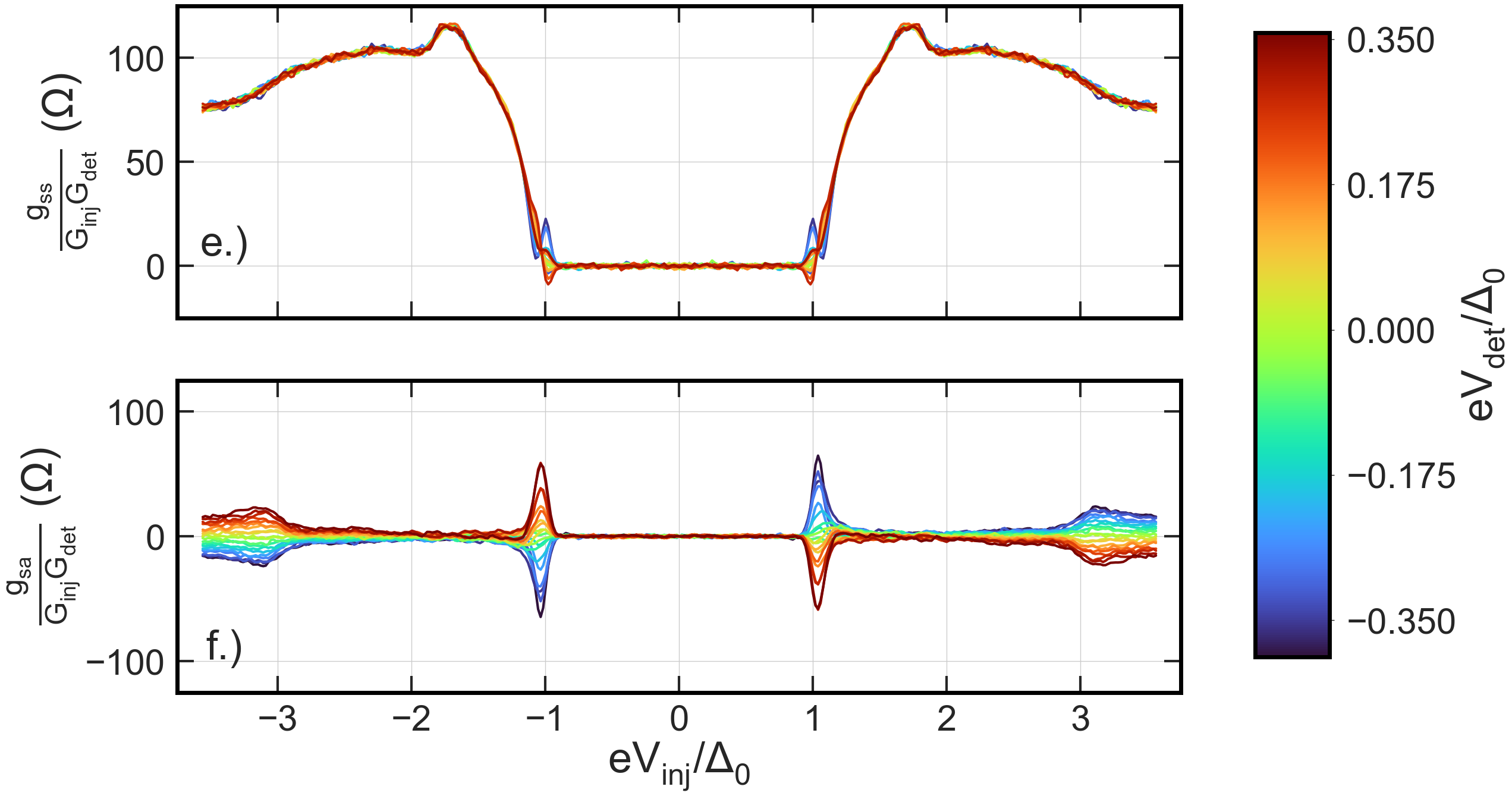}
    \fi
    \caption{ a.)+b.) Normalized non-local conductance between the $A/B$ and $A/C$ junction pairs with a small detector bias. c.)+d.) and e.)+f.) Corresponding decompositions of the non-local dual-biased conductance data into its components symmetric/antisymmetric with respect to $V_\mathrm{inj}$. All color-maps are identical between panels.} %The data has been interpolated for analysis.}
    \label{fig:SubGapBiasDecomp}
\end{figure*}

\subsection{\label{ss:Experimental} Asymmetry of the Non-local Conductance}
To begin, we consider first just the charge imbalance signal that is observed at zero detector bias due $\mu_{Q^*}$. To account for differences in the area for each junction, we divide the measured non-local signal by the normal state conductance of both the injector and detector junctions (denoted by the subscripts) for the purpose of comparison, as shown in Fig.~\ref{fig:ChargeImalanceSignal}. This signal is positive by convention and symmetric with respect to  the injector bias (i.e. particle-hole symmetry), but obeys an exponential relaxation with increased separation between the junctions over a length $\lambda_{Q^*}$. In Ref.~\onlinecite{hubler_charge_2010}, this length was seen to be maximal near $eV/\Delta_0\approx 2$, with the non-local conductance dropping slightly at higher energies, which was explained by increased inelastic scattering at higher quasiparticle energy. In the present device, the junction separation is on the order of $\lambda_{Q^*}$ and $\xi$, and the overall charge imbalance signal drops much more sharply with increasing energy when compared to Ref.~\onlinecite{hubler_charge_2010}, being reduced by approximately 30\% by $eV_\mathrm{inj}/\Delta_0 \approx 3.5$.

We can see this transition in the inelastic relaxation of injected quasiparticles more clearly by biasing the detector slightly away from the Fermi-level. For small biases $eV_\mathrm{det}/\Delta_0 \ll 1$, a charge current is also expected due to normal tunneling by Eq.~\ref{eqn:Inonlocal} from which we can infer information about the energy mode imbalance. This occurs as, when the $f_\mathrm{T}$ mode is excited, small variations in $\Delta$ yield a non-local conductance via Eq.~\ref{eqn:Inonlocal}. This can be clearly seen in Fig.~\ref{fig:SubGapBiasDecomp}, where we demonstrate this effect in both the $A/B$ and $A/C$ injector/detector junction pairs on our device, for detector biases in the sub-gap regime where there remains negligible \textit{local} conductance for the detector.

The key characteristics of these results are the appearance of strong ``spikes" about $eV_\mathrm{inj}/\Delta_0 \approx \pm 1$, and the onset of further anti-symmetric components for $|eV_\mathrm{inj}/\Delta_0| \gtrapprox 3$. Both effects also respect the same anti-symmetry with the sign of the detector bias $V_\mathrm{det}$. As these features appear strongly during injection at the gap edge (where the available quasiparticle states are mainly charge neutral excitations) it is reasonable to expect that all of the observed anti-symmetric conductance here is associated with energy imbalance. Given this, and the corresponding reduction in charge imbalance at these energies, the reappearance of the anti-symmetric non-local conductance above $|eV_\mathrm{inj}/\Delta_0| \gtrapprox 3$ indicates a transition in the inelastic relaxation rate of fully charged quasiparticles into charge neutral excitations. This could be expected to occur due to secondary pair-breaking via electron-electron or electron-phonons scattering which is sometimes referred to as quasiparticle multiplication~\cite{burnell_inelastic_1994}. In the $A/C$ junction pair, there is additionally a slight ``hump" near $|eV_\mathrm{inj}/\Delta_0| \gtrapprox 1.75$. This is similar in appearance to features sometimes seen in the local conductance of these junctions, and could be a convolution of the phonon spectra to the tunneling conductance.% which affects the generation efficiency of charge imbalance at certain energies.

\begin{figure}[t!]
    \centering
    \ifdefined\withfigures
    \includegraphics[width=0.95\linewidth]{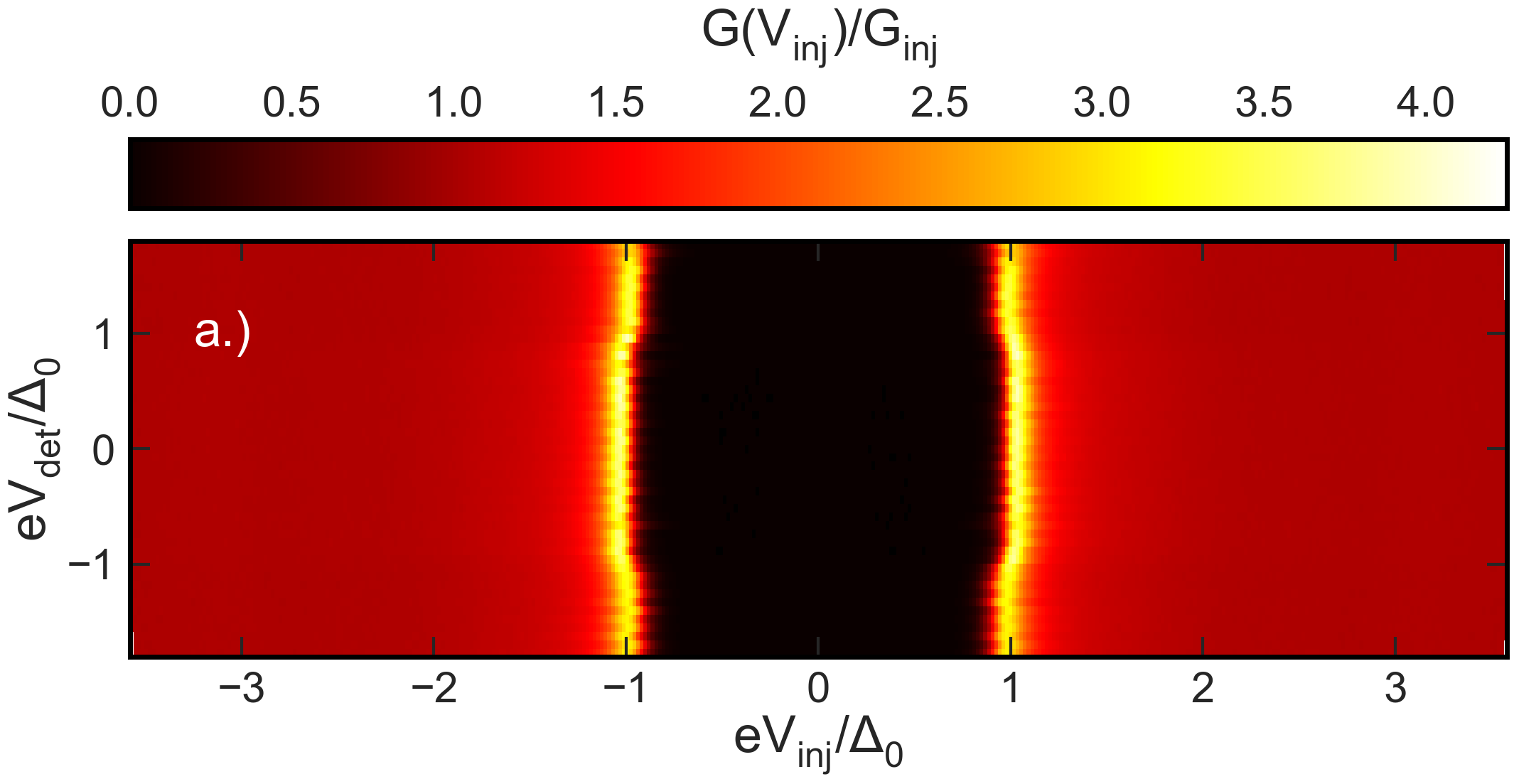}\\
    \includegraphics[width=0.95\linewidth]{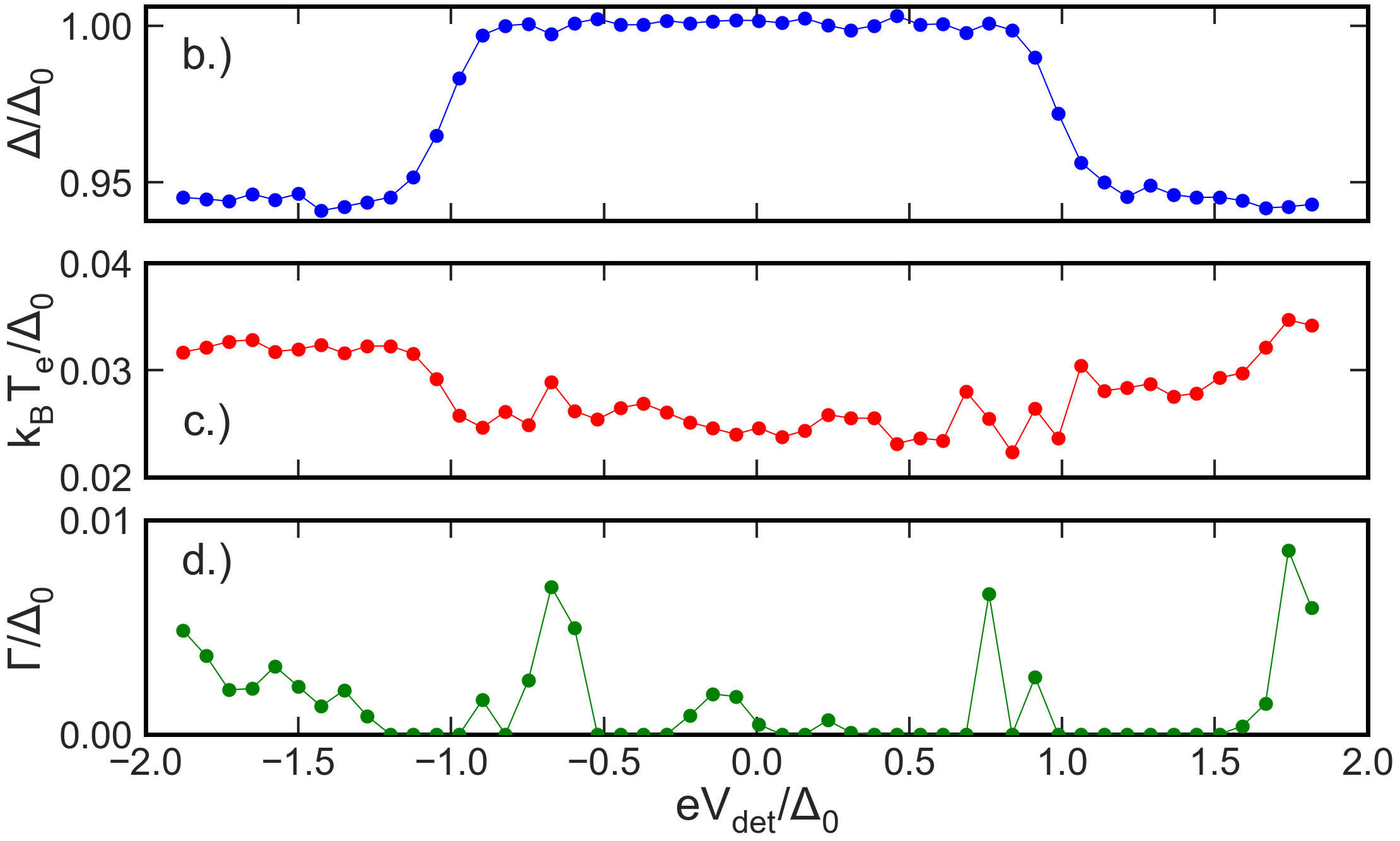}
    \fi
    \caption{a.) Heatmap plot of the local conductance of the injector junction $A$ during simultaneous detector biasing on $B$. b.)+c.)+d.) Results of fittings to a Dynes-type model for the DOS, plotted as a function of detector bias. For all fits, only the normal state conductance was held fixed, with the other parameters varying freely.}
    \label{fig:GapSupression}
\end{figure}
As the detector bias is increased further towards the equilibrium gap edge, there is a reciprocal effect onto the non-equilibrium gap seen at the injector due to non-equilibrium quasiparticles there also. This can be seen clearly in the conductance of the injector junction during simultaneous biasing as shown in Fig.~\ref{fig:GapSupression} for the $A/B$ pair. There, the non-equilibrium gap at the injector junction is sharply suppressed by the back-action of the detector, causing a shift in the local conductance spectra corresponding to $\approx6\%$ reduction from $\Delta_0$. Such a contribution occurs due to the self-consistency relation for $\Delta$ and the large DOS at the BCS gap edge. Note that the detector bias was not seen to cause a large corresponding increase in the Dynes parameter, nor in the effective electronic temperature of the injector junction, as was reported in Ref.~\onlinecite{kolenda_nonlocal_2013} during a similar experiment. There, the authors argued that a large jump in the current due to Joule heating of the normal metal counter-electrodes during detector biasing dominated the quasiparticle transport by shifting the temperature equilibrium of the superconductor indirectly. Here, we see little evidence for this occurring due to the much lower lead resistance and normal state tunnel-conductance of the present devices, and the gap suppression is due directly to the injection of non-equilibrium quasiparticles by the detector. %Consequently, we can be fairly certain that the asymmetric gap-edge spikes in the non-local conductance data are due to injected charge neutral quasiparticles tunneling out of superconductor.

\begin{figure}[t!]
    \centering
    \ifdefined\withfigures
    \includegraphics[width=0.95\linewidth]{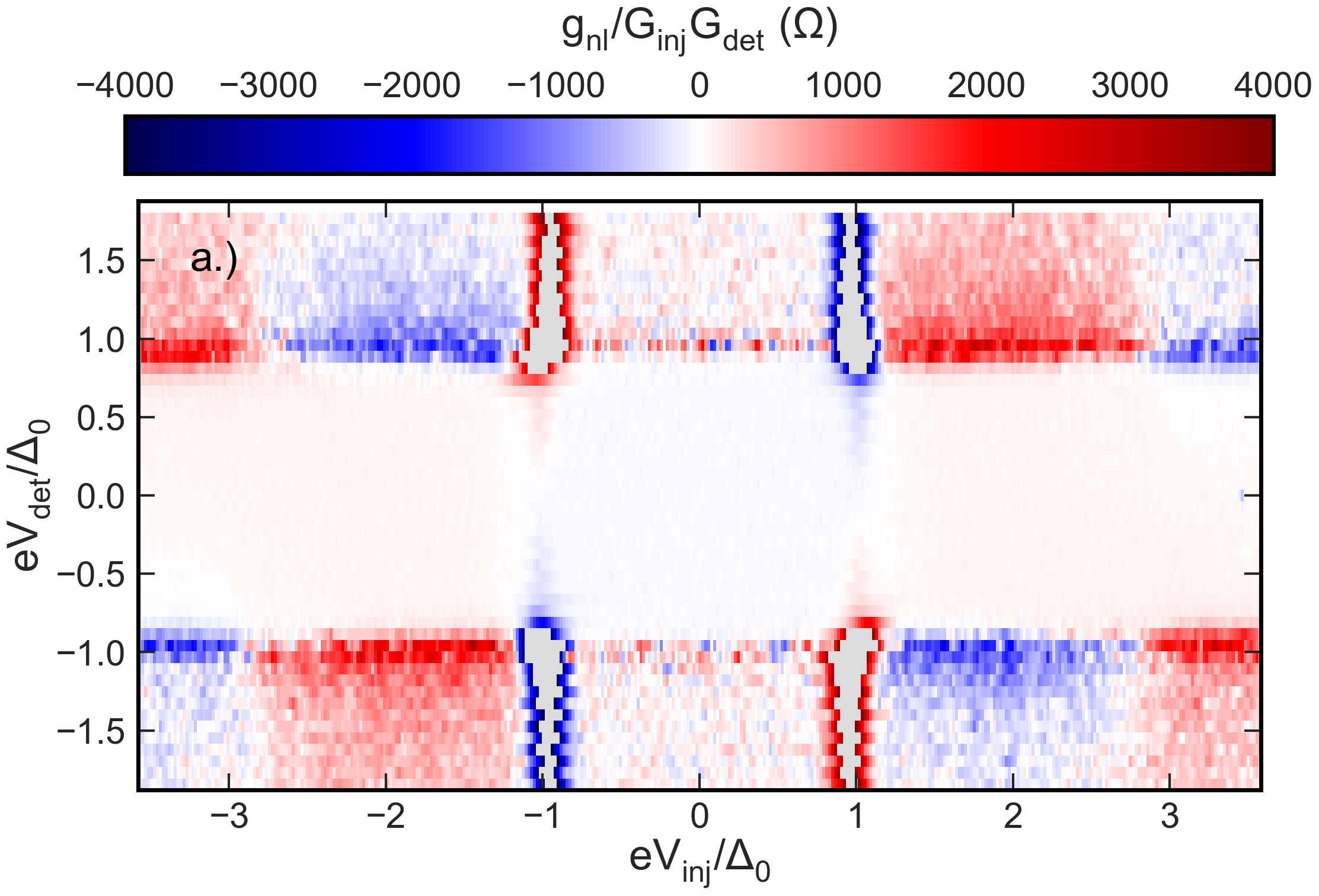}\\
    \includegraphics[width=0.95\linewidth]{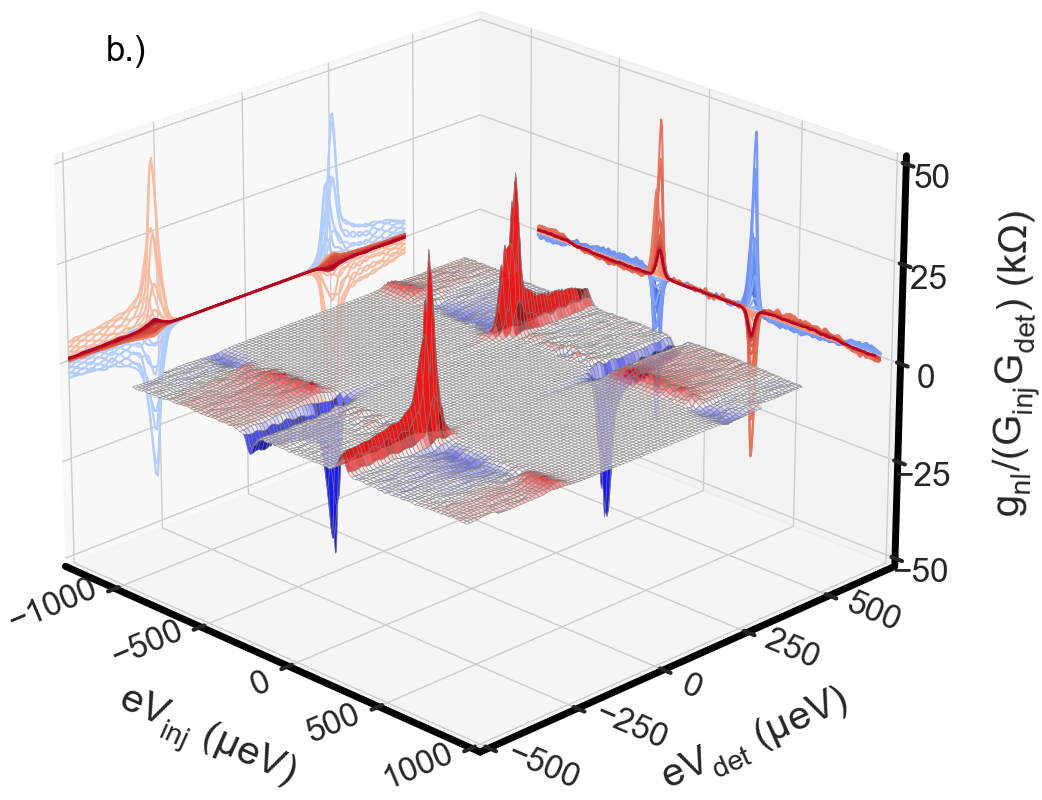}
    \fi
    \caption{a.) Heatmap of the dual-biased non-local conductance data for the $A/B$ junction pair. The grey regions correspond to data outside the color-map. b.) Same data, presented as a 3D contour over its full data range. Also shown are projections indicating the general shape of the data taken during biasing of either $V_\mathrm{inj}$ or $V_\mathrm{det}$, indicating the sign reversal of the signal with respect to the bias polarity.}
    \label{fig:FullHeatmaps}
\end{figure}

Considering again the non-local conductance measured in this same bias range, there is now a significant spectral density of quasiparticles at the energy of the detector which greatly increases the magnitude of the current and resulting signal. In Fig.~\ref{fig:FullHeatmaps}.a, we report this non-local signal measured simultaneously with the previous local data, over a limited range chosen to highlight the dominant anti-symmetric response for large detector biases. This shows a clear extension of the spikes at the injector gap edge $|eV_\mathrm{inj}|\approx\Delta_0$ up into the region where $|eV_\mathrm{det}|\gtrapprox \Delta$. In Fig.~\ref{fig:FullHeatmaps}.b, we plot the same dataset fully as a 3D contour, where one can see clearly that these spikes are located around $(eV_\mathrm{inj}, eV_\mathrm{det}) \approx \pm\Delta_0$ are are over $400\times $ greater in magnitude than the charge imbalance contribution considered before in Fig.~\ref{fig:ChargeImalanceSignal} and Fig.~\ref{fig:SubGapBiasDecomp}. At higher injector biases we consistently find a reversal of the anti-symmetric non-local conductance which is also many times larger than the charge imbalance contribution. We have seen a similar non-local conductance previously during dual bias of larger devices in which the injector and detector spacing was comparable to $\lambda_{Q^*}$, as reported in Ref.~\onlinecite{ryan_enhanced_2025}, and in several other test samples (not reported). In this representation, the effect generally appears as antisymmetric lobes in the conductance over a mostly symmetric background.  As we will discuss later in our theoretical results, this sign reversal is inconsistent with existing models of quasiparticle transport, leading us to suspect it to be a true non-equilibrium transport effect due to the simultaneous injection of quasiparticles into both junctions. 

\subsection{\label{ss:InjectedCurrent} Non-local Conductance at Large Supercurrent Bias}

As we have seen so far, NIS junctions can be made sensitive to the energy-mode imbalance via the gap-suppression caused by a charge neutral quasiparticle distribution. We can use this effect to observe how quasiparticles re-distribute in the presence of external variables, such as a magnetic field (see Appendix~\ref{App:MagField}) or a large supercurrent injected across the wire. As our devices are constructed from an un-shunted \ce{Al} wire, we are able to perform the latter test by using large series resistors on either side of the wire, and a floating voltage source to generate a current across it. As nominally no current is being carried in the normal metal layer (which is also isolated by a tunnel barrier from the \ce{Al}) this results in a pure supercurrent coming from the superconducting reservoirs. This current is then drained via an additional low-impedance connection to ground out the left side of the wire, which maintains the potential of the superconducting wire near zero despite the large current and helps prevent a spurious voltage from being developed across the junctions. To bias the device into the regime of interest, a small fixed offset voltage was applied to the detector junction ($eV_\mathrm{det} \approx \SI{30}{\micro eV}$) and either the injector or supercurrent were swept. Just below the critical current of a wire one expects depairing to occur by Eq.~\ref{eqn:depairing} leading to an increase in the overall relaxation and recombination rates for quasiparticles. As discussed earlier, the generated phase gradient also couples the $f_L$ and $f_T$ distributions, which will further perturb the non-local conductance spectra.

\begin{figure}[tb!]
    \centering
    \ifdefined\withfigures
    \includegraphics[width=0.95\linewidth]{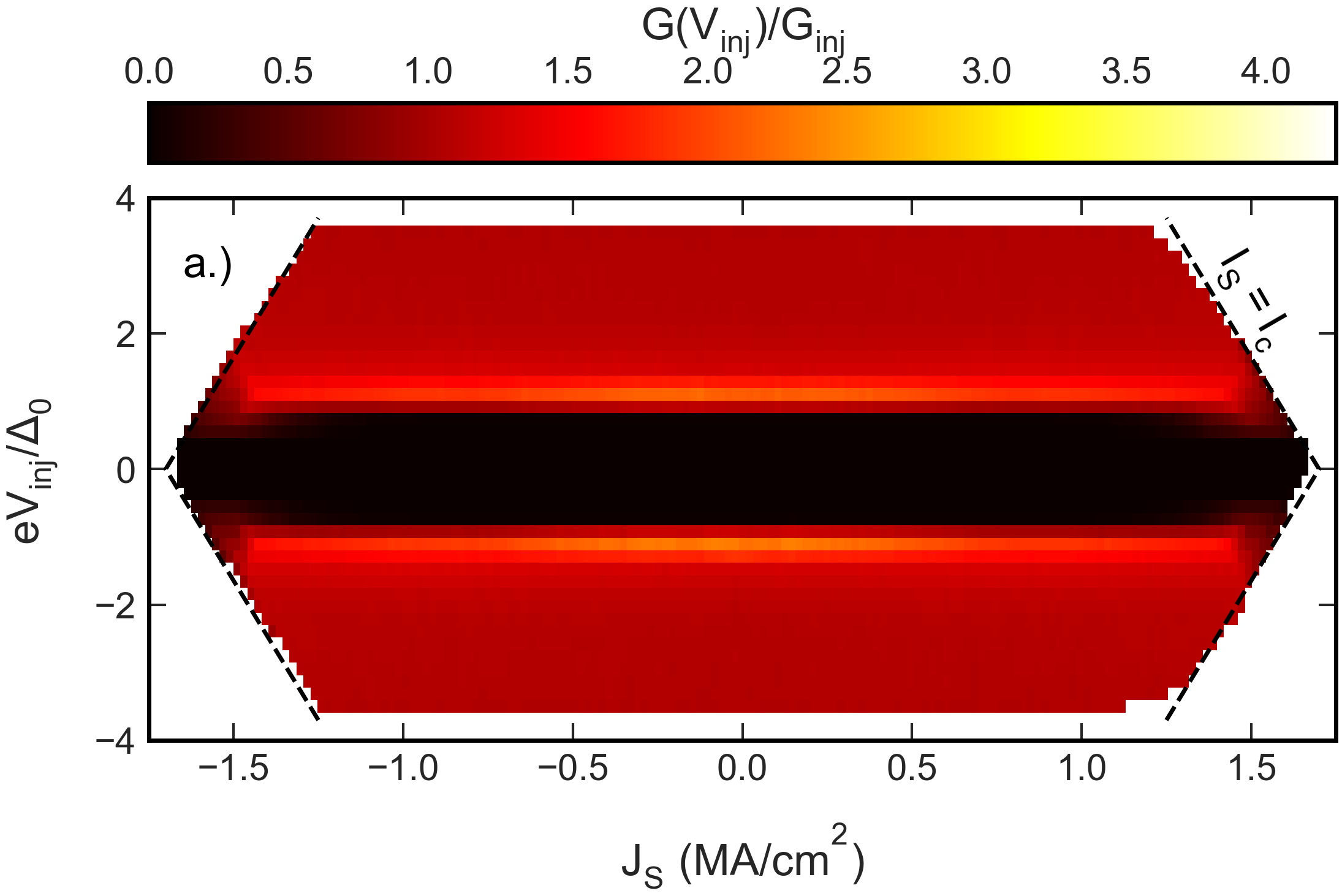}\\
    \includegraphics[width=0.95\linewidth]{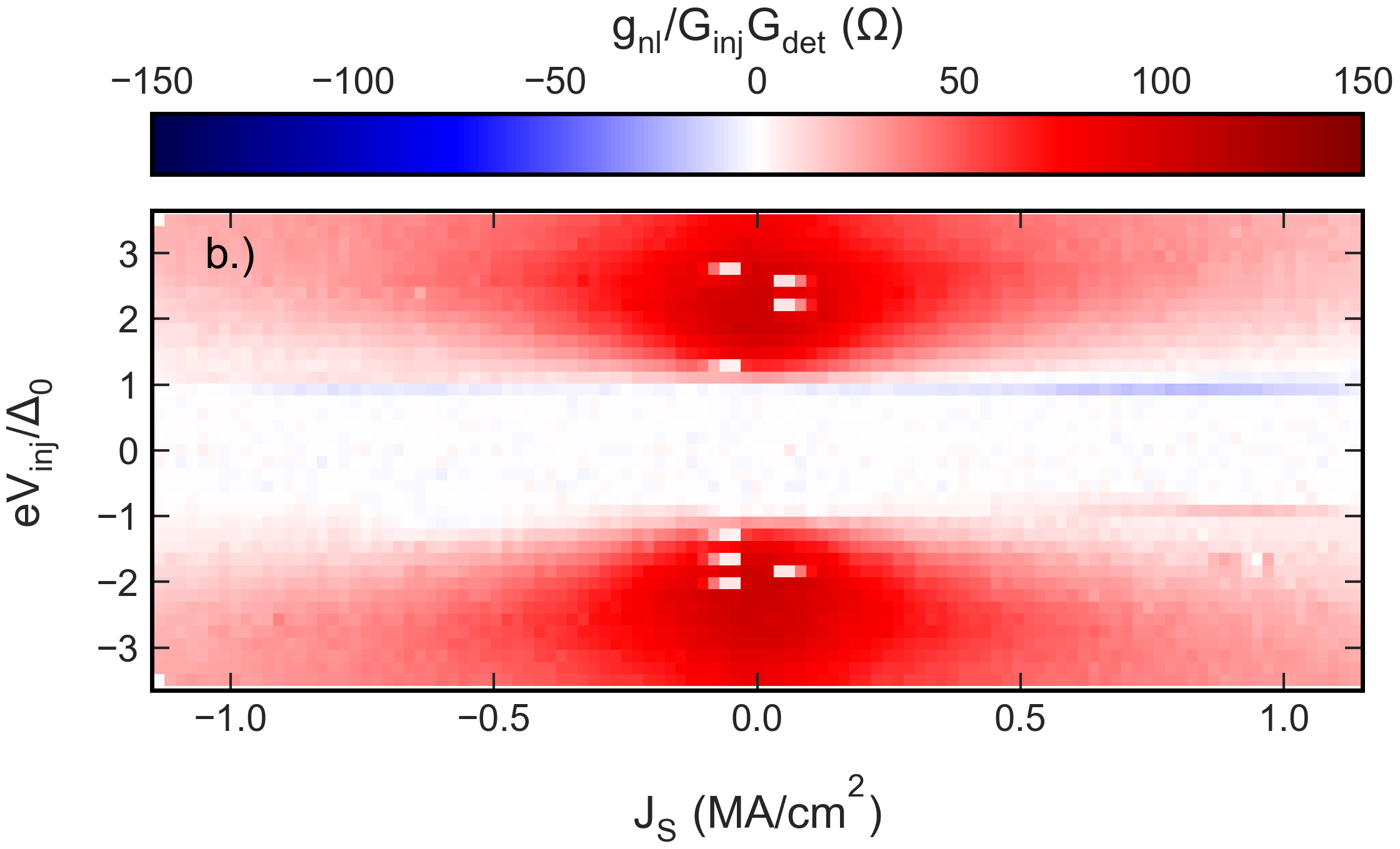}    
    \fi
    \caption{a.) Heatmap of the local conductance of the injector junction as a function of applied supercurrent density and injector bias. The dashed black lines indicate the point where the total supercurrent exceeds the critical current during the sweep, and the data has been suppressed to ignore parts of the sweep beyond this where the device is driven normal. b.) Non-local conductance heatmap measured between the $A/B$ pair over a narrower range of $J_S$ values under swept current bias. Only a slight indication of the non-local peaks are discernible. The artifacts near zero current bias are caused by stochastic switching and retrapping which has been interpolated out of the dataset wherever possible. For estimation of $J_S$, a cross sectional area of \SI{1.2E3}{\nano\meter\squared} was assumed, corresponding to the total un-oxidized volume of Al, and should thus be taken as a slight underestimation.}
    \label{fig:Supercurrent_LocalEffects}
\end{figure}

To see this, we first sweep the applied supercurrent all the way through the transition into the normal state above the critical current $I_c$ as shown in Fig.~\ref{fig:Supercurrent_LocalEffects}.a. By tracing this out with respect to the injector voltage, we see that there is a linear reduction of $I_c$. As the critical current of the wire is on the order of $10^3$ times the maximum amplitude of the quasiparticle current injected, this is not simply due to the combination of currents, and is likely caused by the energy imbalance due to quasiparticle injection. As $I_s$ is increased, this leads to a rapid suppression of the pair potential, followed by a switch into the normal state for the entire wire. This is even seen below the gap for $|J_S| \geq\SI{1.5}{M\ampere/\centi\meter\squared}$, where a finite conductance was observed at the injector junction while the wire remained at zero resistance.

We next consider the non-local effects, where we have repeated this measurement over a narrower bias range in order to minimize stochastic switching artifacts in the data. As seen in Fig.~\ref{fig:Supercurrent_LocalEffects}.b, the dominant charge imbalance signal now appears as a pair of lobes, which extend to biases $|eV_\mathrm{inj}/\Delta_0|\gtrapprox 2$ and diminish with the magnitude of the applied $J_S$. This can be understood in terms of the increased charge relaxation rate due to the depairing effect of the supercurrent (see the fits in Fig.~\ref{fig:fit_fT_vs_Is}). 

\begin{figure}[tb!]
    \centering
    \ifdefined\withfigures
    \includegraphics[width=0.95\linewidth]{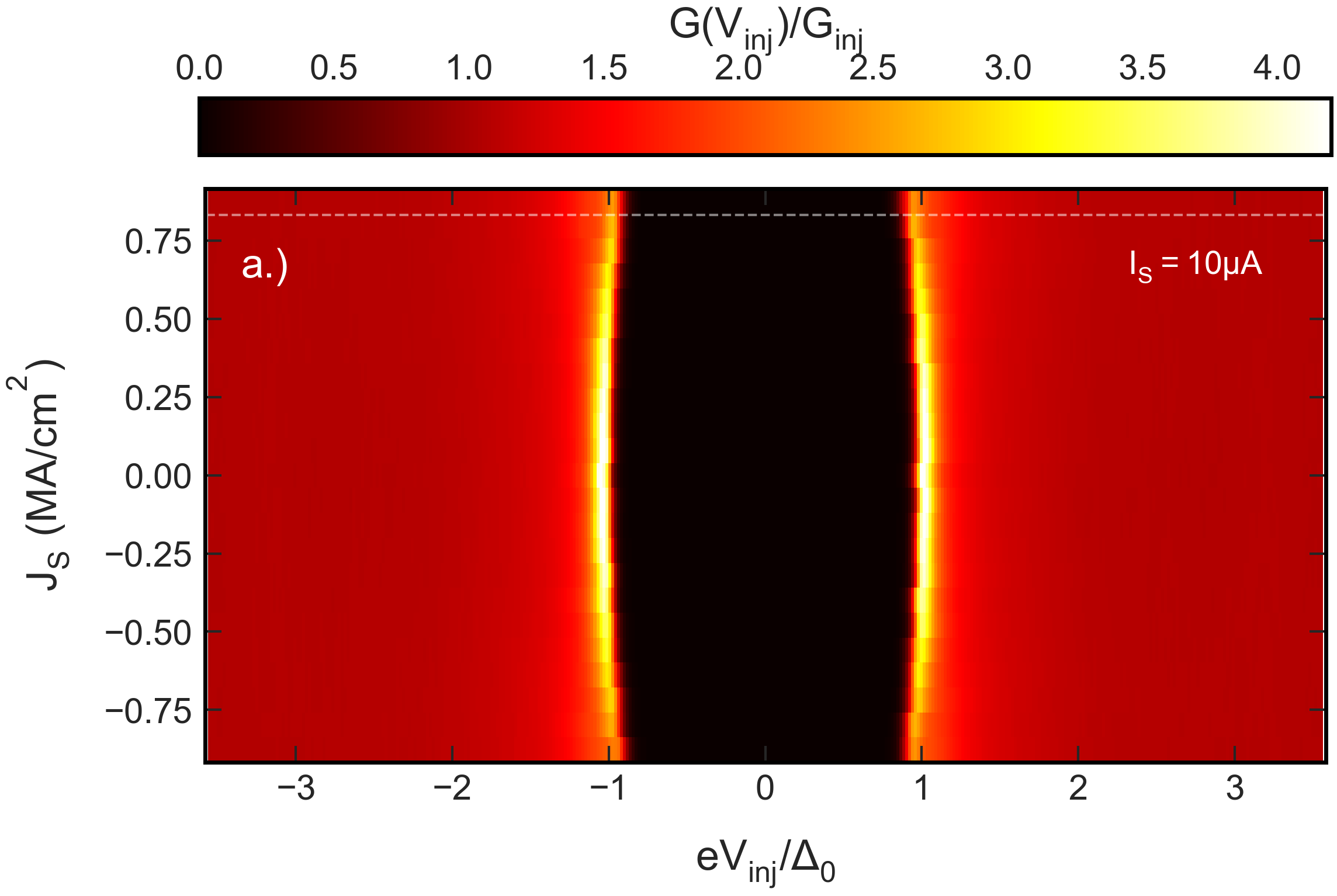}\\
    \vspace{2em}
    \includegraphics[width=0.95\linewidth]{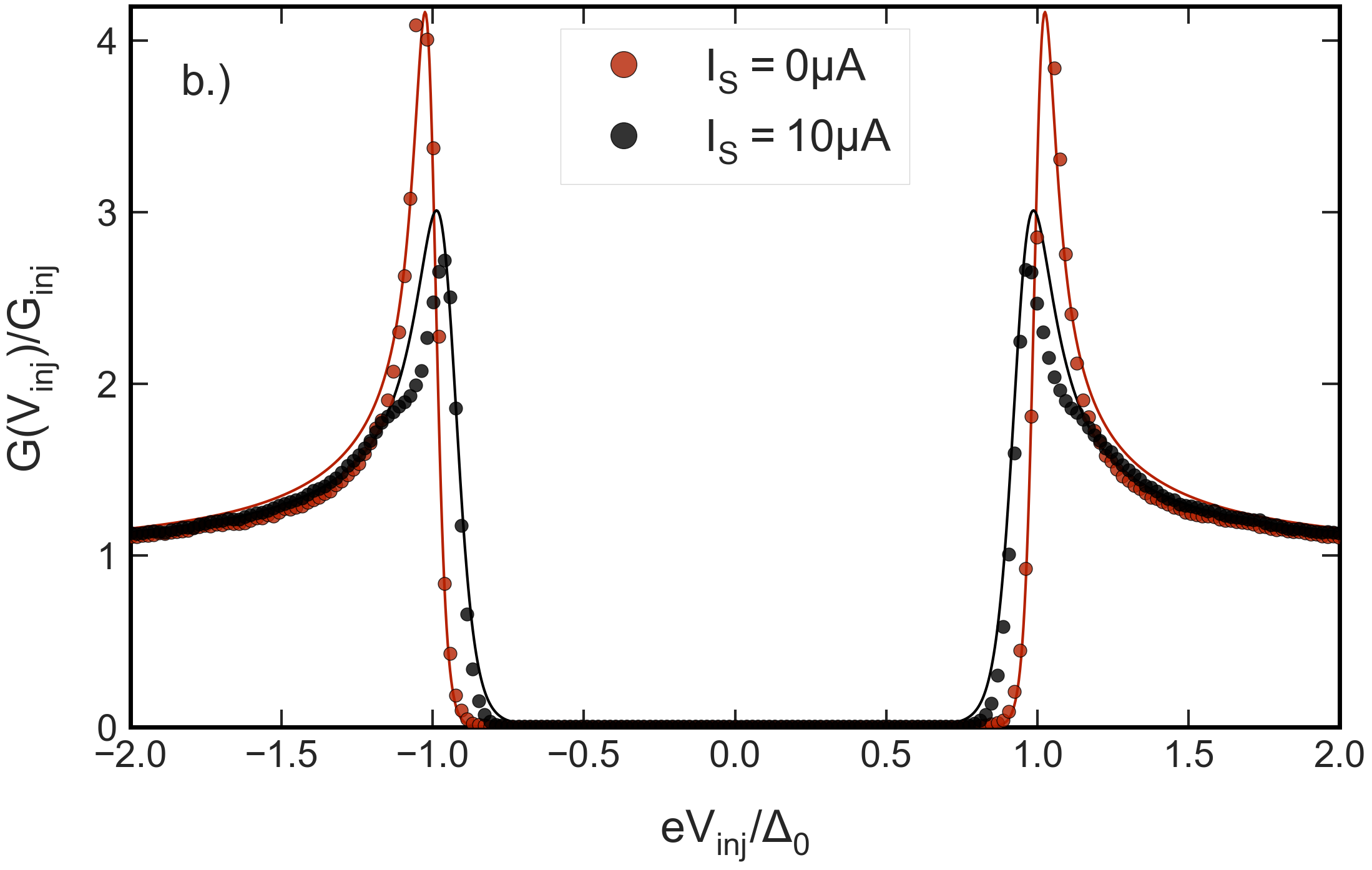}    
    \fi
    \caption{a.) Heatmap of the local conductance spectra measured on the $A$ junction as a function at fixed supercurrent bias. The total current on the y-axis corresponds to a range $I_S =\pm \SI{11}{\micro \ampere}$. b.) Line traces taken at 0 and \SI{10}{\micro\ampere} from the same measurement, demonstrating the non-equilibrium conductance spectra which occur with increasing supercurrent bias. }
    \label{fig:NonBCSDOS}
\end{figure}

\begin{figure}[tb!]
    \centering
    \ifdefined\withfigures
    \includegraphics[width=0.95\linewidth]{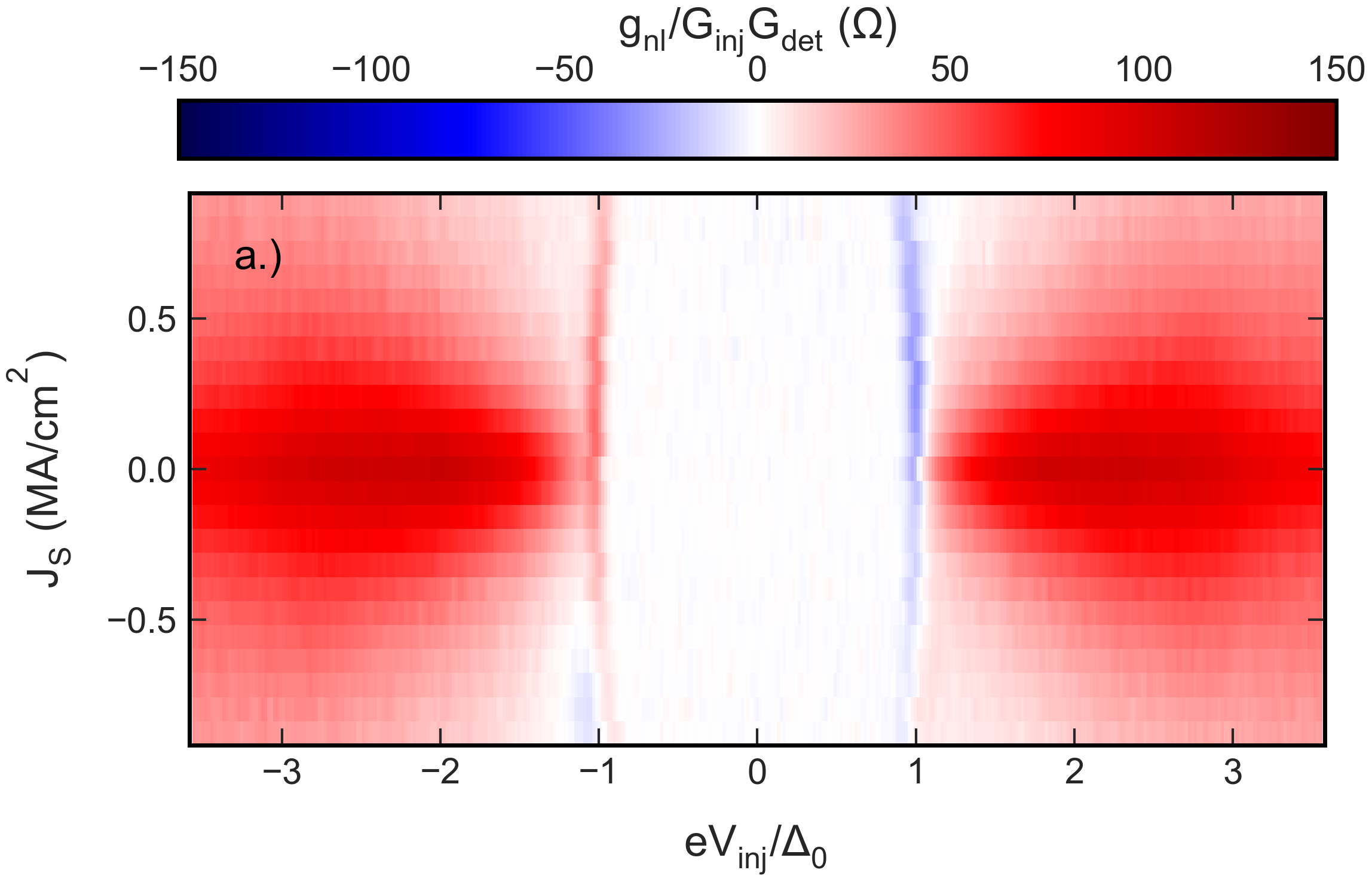}\\
    \vspace{1em}
    \includegraphics[width=0.95\linewidth]{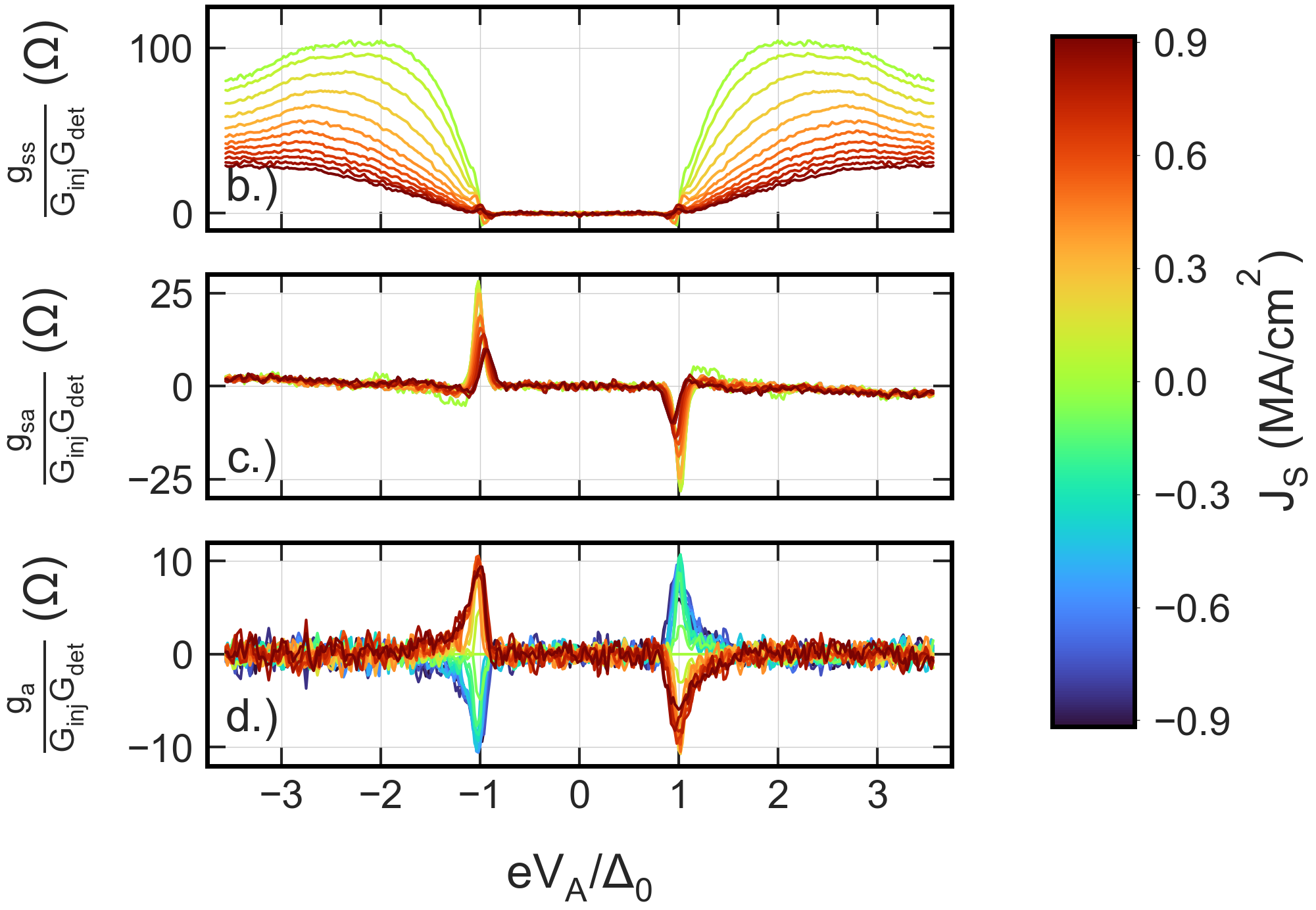}    
    \fi
    \caption{a.) Heatmap of the non-local conductance spectra measured simultaneously with the data in Fig.~\ref{fig:NonBCSDOS} on the $A/B$ junction pair. b.)+c.) Corresponding decompositions of the non-local conductance data into its components symmetric/antisymmetric with respect to the injector voltage $V_\mathrm{inj}$ for each value of the injected supercurrent.}
    \label{fig:NonLocSuperBiasHeatMap}
\end{figure}

If we instead sweep the injector voltage at fixed supercurrent, we find that we can directly observe the resulting quasiparticle spectra as well as trace out the local and non-local conductances based on the symmetry considerations discussed before. For instance, gap suppression due to pair-breaking can be seen in the inward bowing of the coherence maxima shown in Fig.~\ref{fig:NonBCSDOS}.a, while for $I_S \gtrapprox I_c/2$ one begins to see additional symmetric shoulders appearing at slightly higher energies, which we demonstrate in Fig.~\ref{fig:NonBCSDOS}.b, clearly demonstrating the non-equilibrium/non-thermal state of the wire. Given the geometry of the device, it is possible that this may be in essence a proximity effect due to the superconducting reservoirs themselves, where the gap will maintain its equilibrium value as the current is injected. In the non-local conductance, we see both a suppression in the $V_\mathrm{inj}$ symmetric component independent of the sign of $I_S$, as well as a broadening and sign reversal of the gap-edge peak which appears to depend on both the magnitude and the sign of the supercurrent (i.e., on the sign of $\nabla \phi$). This is most apparent in Fig.~\ref{fig:NonLocSuperBiasHeatMap}.a for $eV_\mathrm{inj}/\Delta_0 \approx -1$ for large negative values of $J_S$. As the local conductance spectra remain symmetric even at large $I_S$, one is left to conclude that the $f_L$ and $f_T$ modes are being mixed per the kinetic equations.

\section{\label{s:Discussion}Discussion}

% \subsection{\label{ss:Comparison}Comparison to Theory}

\begin{figure}[tb!]
	\centering
    \ifdefined\withfigures
	\includegraphics[width=0.95\linewidth]{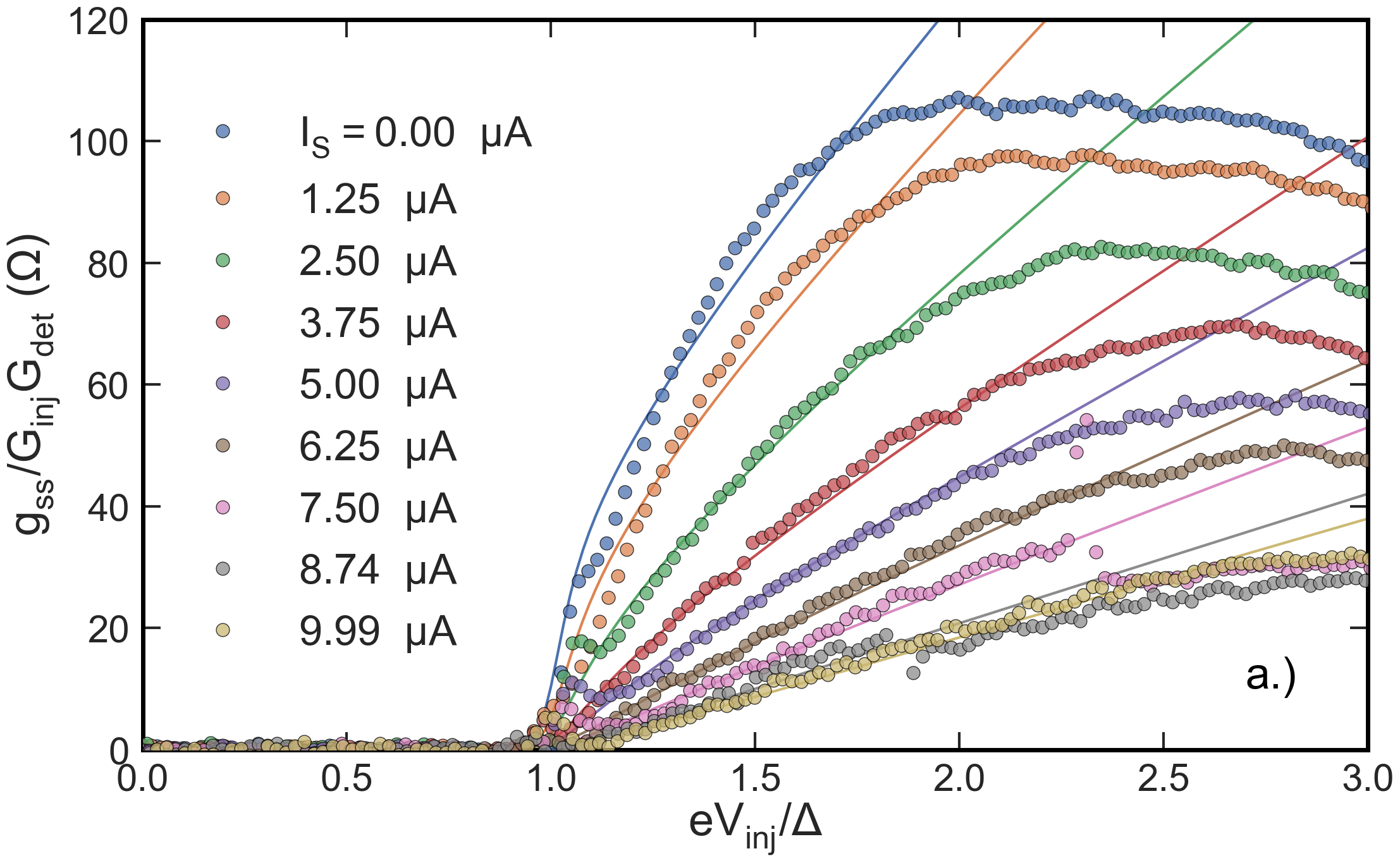}\\
    \includegraphics[width=0.95\linewidth]{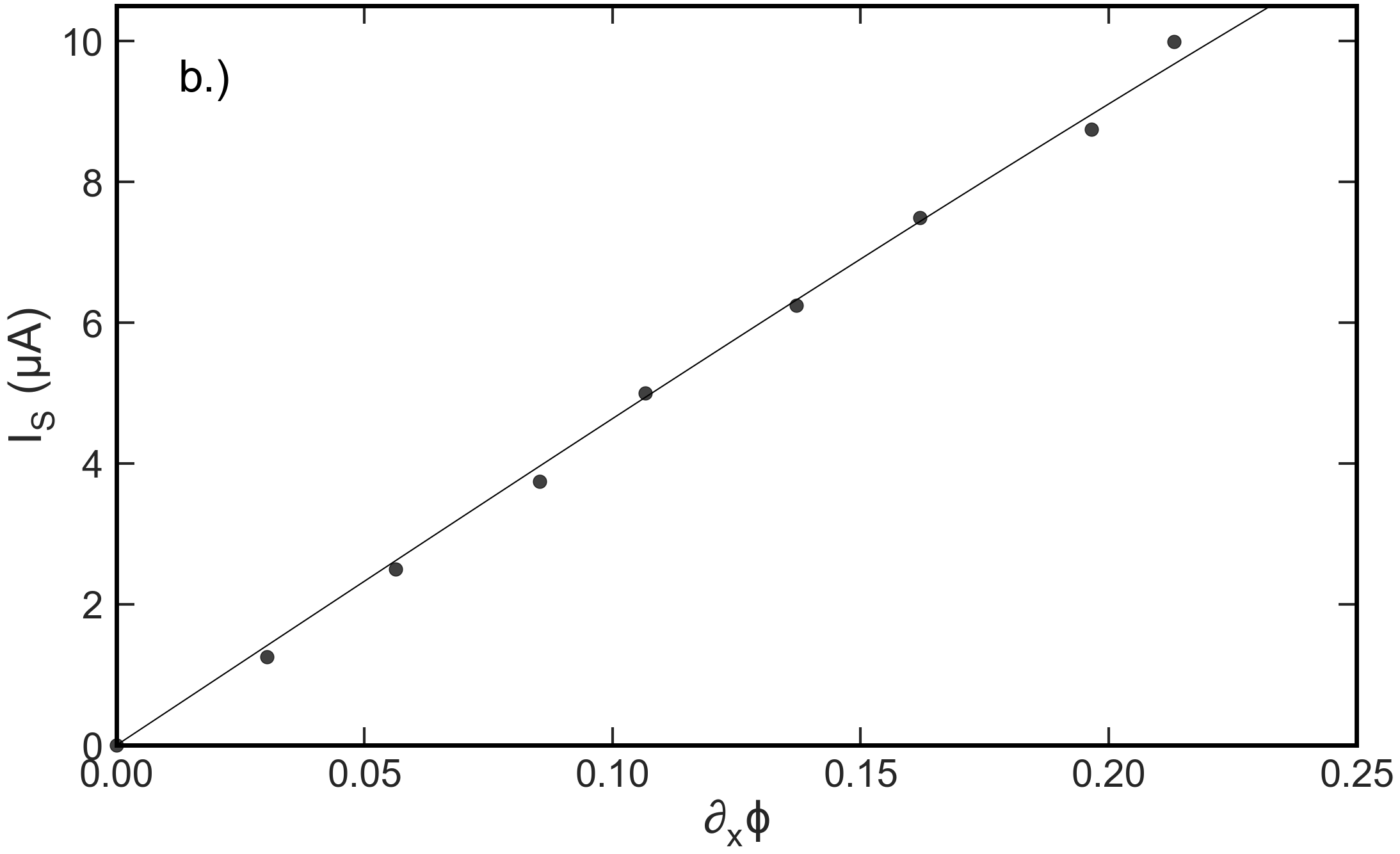}
    \fi
	\caption{a.) Symmetric part $g_\mathrm{ss}$ of the nonlocal conductance as a function of bias for different supercurrents $I_\mathrm{s}$. Symbols are data from Fig.~\ref{fig:NonLocSuperBiasHeatMap}.b, lines are fits (see text). b.) Applied supercurrent $I_\mathrm{s}$  as a function of phase gradient $\partial_x\mop$ obtained from the fits in a, with a fit to Eq.~(\ref{eqn:is}).}
	\label{fig:fit_fT_vs_Is}
\end{figure}

To demonstrate the mechanisms which give rise to the asymmetric nonlocal conductance, we next discuss the corresponding components of the conductance via quasiclassical simulations of the device. We first consider the charge imbalance component symmetric in both bias and current, $g_\mathrm{ss}$, as shown in Fig.~\ref{fig:fit_fT_vs_Is}.a as a function of bias for different supercurrents $I_\mathrm{s}$ along the wire. The lines are fits of the data using the analytic solution of the kinetic equation in the absence of inelastic scattering (and neglecting the coupling term). Only the initial increase of the signal for $eV_\mathrm{inj}/\Delta<1.5$ is included in the fit. For higher bias, inelastic scattering leads to a downward deviation from the fit \cite{hubler_charge_2010,wolf2014}. For $I_\mathrm{s}=0$, the lifetime parameter $\Gamma=(1.53\pm0.02)\times10^{-3}$ is the only fit parameter. For $I_\mathrm{s}\neq0$, $\Gamma$ is kept constant, and the phase gradient $\partial_x \mop$ is the only fit parameter.  Fig.~\ref{fig:fit_fT_vs_Is}(b) shows $I_\mathrm{s}$ as a function $\partial_x \mop$ obtained from the fit, together with a fit to Eq.~(\ref{eqn:is}), with $I_0=29.8\pm0.3~\mathrm{\mu A}$ as the only fit parameter, slightly smaller than the estimate $I_0=30.9~\mathrm{\mu A}$ from the sample parameters. As such, the overall suppression in the magnitude of the charge imbalance signal under an external supercurrent can be understood in terms of the depairing contribution due to the nonzero $\partial_x\mop$ per Eq.~\ref{eqn:depairing}. 

\begin{figure}[tb!]
	\centering
    \ifdefined\withfigures
	\includegraphics[width=0.95\linewidth]{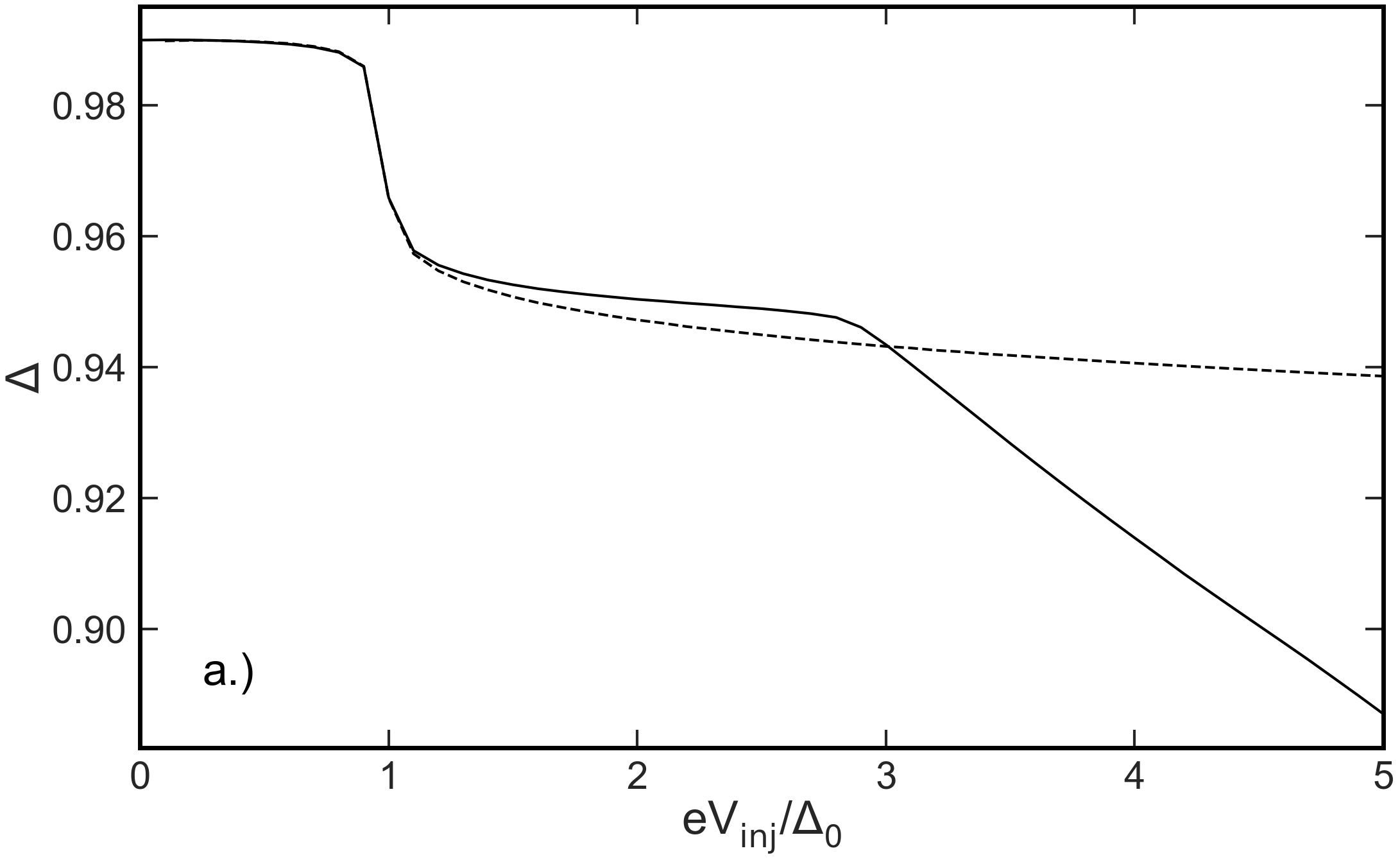}\\
    \includegraphics[width=0.95\linewidth]{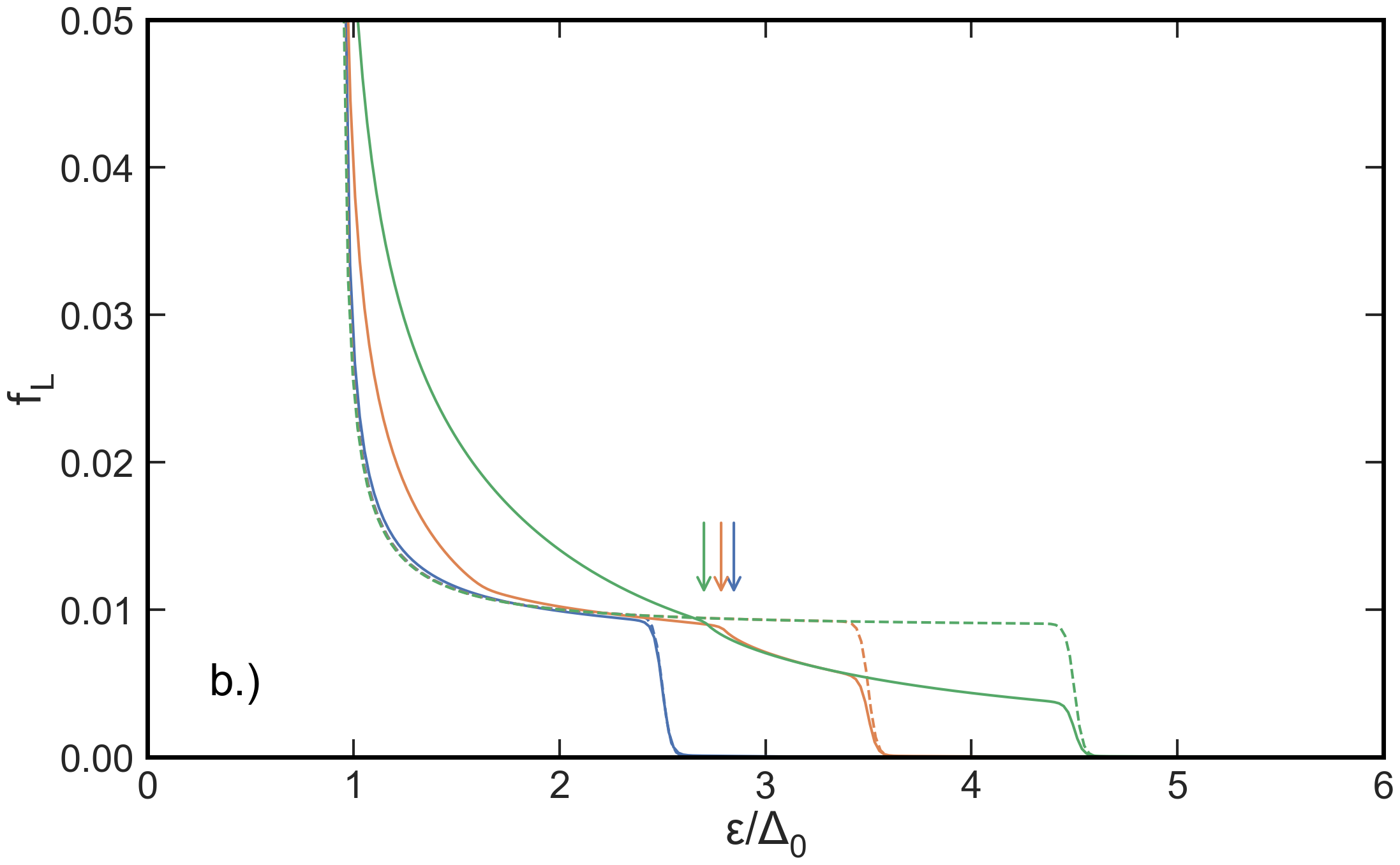}
    \fi
	\caption{a.) Normalized self-consistent pair potential $\Delta$ as a function of normalized injector bias $eV_\mathrm{inj}/\Delta_0$. b.) Distribution function $f_\mathrm{L}$ as a function of normalized energy $\epsilon/\Delta_0$ for different injector bias. The arrows indicate $3\Delta$ with the self-consistent $\Delta$ for the bias point.}
	\label{fig:sim_ee_delta}
\end{figure}

The contribution $g_\mathrm{sa}$, can be understood by simulating the change of $\Delta$ as a function of injector bias. We do this by solving the kinetic equation for $f_\mathrm{L}$, inserting the distribution function at $x=0$ into Eq.~(\ref{eqn:sc}), and iterating until self-consistency is obtained. Strictly speaking, under non-equilibrium conditions $|\Delta|$ should be a function of $x$, while Eq.~(\ref{eqn:kinfL}) assumes a constant $|\Delta|$ along the wire. Since the junction distance is small compared to the length of the wire, we neglect this dependence here to keep the numerical effort tractable. Fig.~\ref{fig:sim_ee_delta}(a) shows the self-consistent pair potential $\Delta$ as a function of injector bias $V_\mathrm{inj}$. Neglecting inelastic scattering (dashed line), the pair potential drops quickly with the onset of quasiparticle injection above $\Delta_0$ as the $f_\mathrm{L}$ is perturbed, and then remains nearly constant. We can then compare this to the case where we include inelastic scattering via the electron-electron collision integral for the $f_\mathrm{L}$ mode given by \cite{bobkova2016}, with a characteristic scattering rate $\gamma_\mathrm{ee}=\hbar/\Delta_0\tau_\mathrm{ee}$. Here, $\tau_\mathrm{ee}$ is the equilibrium inelastic scattering time in the normal state at the Fermi surface at $T_\mathrm{c}$, and we have taken $\gamma_\mathrm{ee}=2\cdot 10^{-4}$ corresponding to $\tau_\mathrm{ee}\approx 12~\mathrm{ns}$.

\begin{figure}[b!]
	\centering
    \ifdefined\withfigures
	\includegraphics[width=0.95\linewidth]{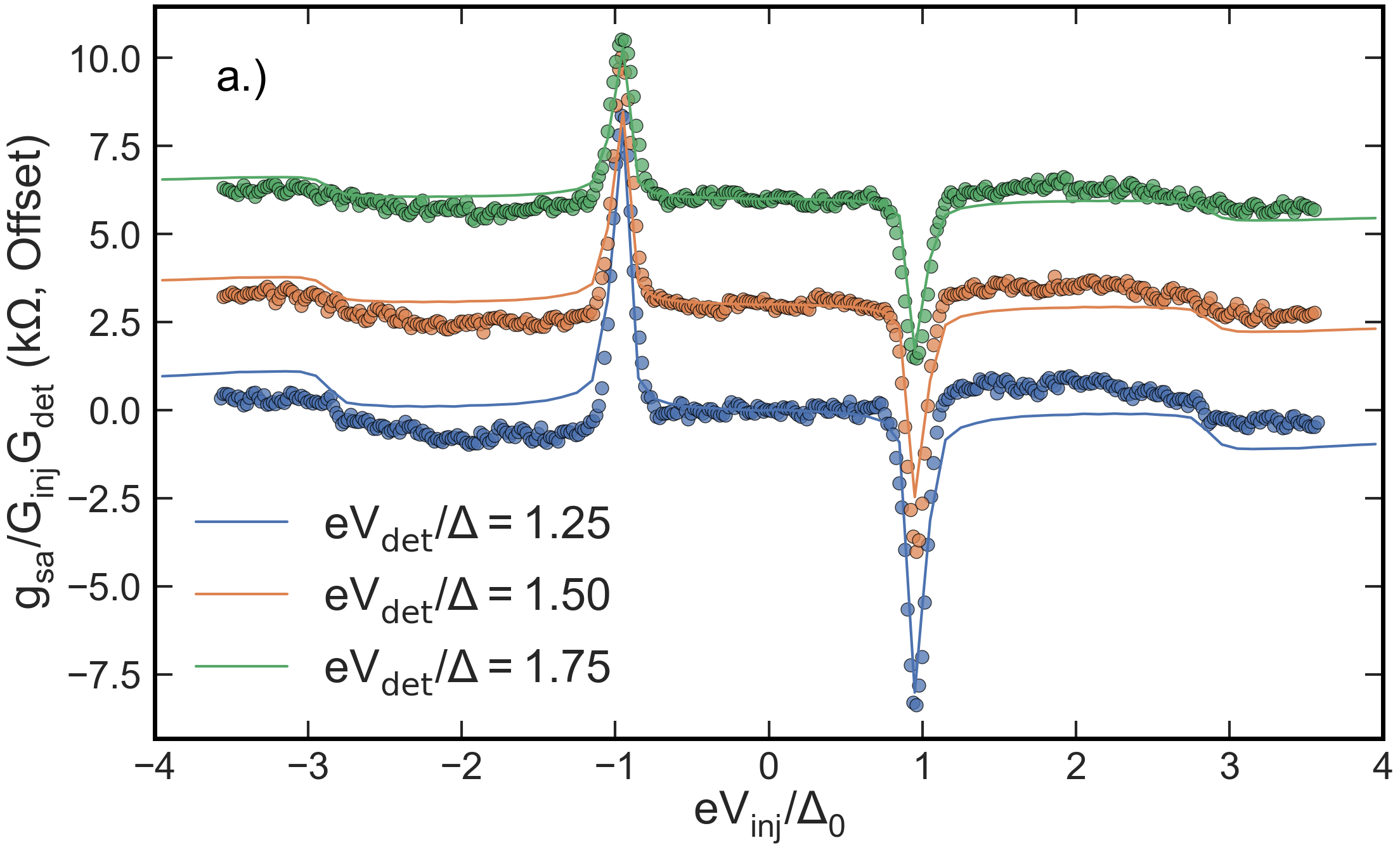}\\
    \includegraphics[width=0.95\linewidth]{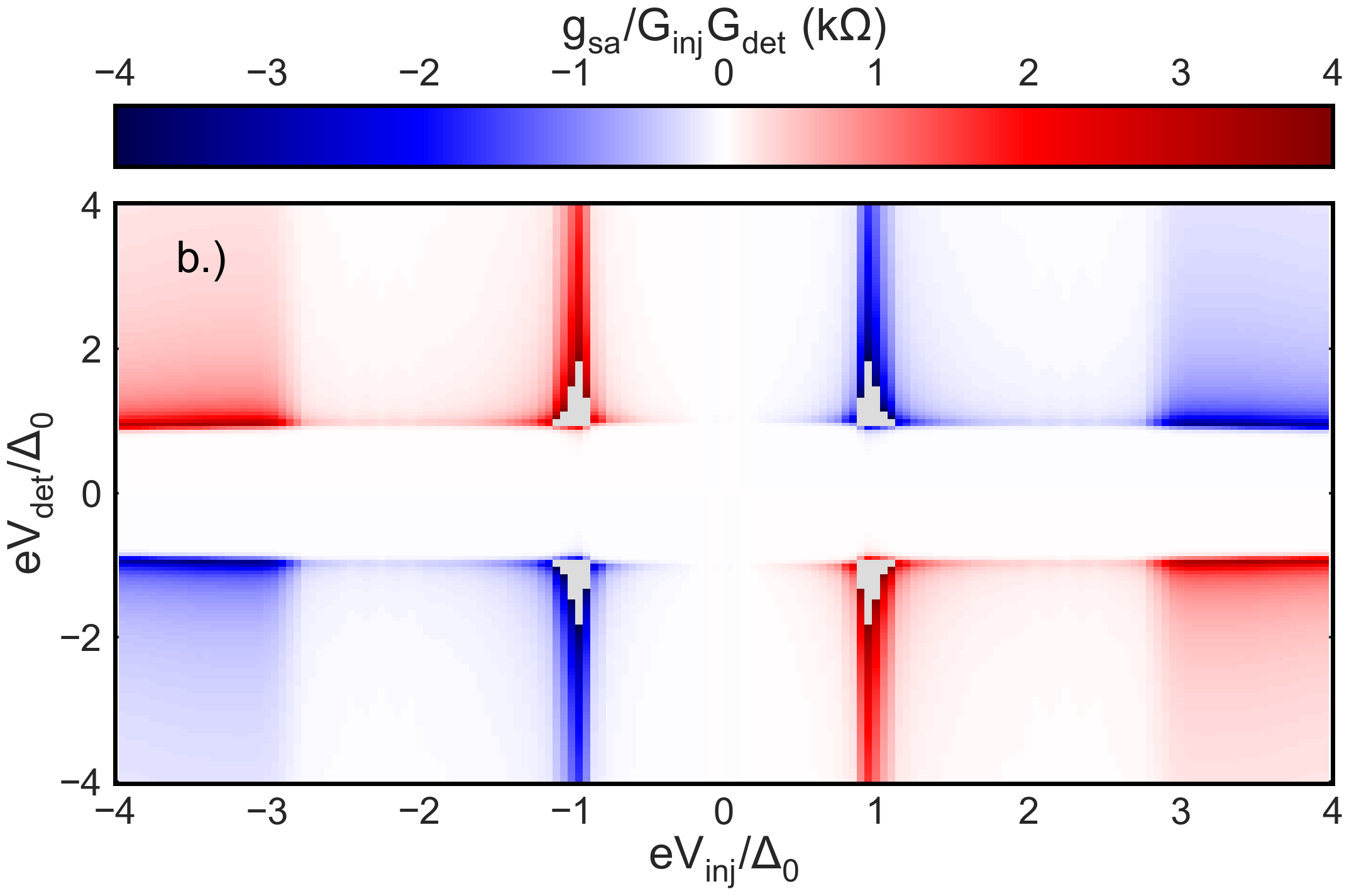}
    \fi
	\caption{a.) Asymmetric part $g_\mathrm{sa}$ of the non-local conductance as a function of injector bias $V_\mathrm{inj}$ for different detector biases $V_\mathrm{det}$. Symbols are experimental data from the $A/B$ junction pair, lines are numerical simulations. The data are offset successively by $3~\mathrm{k\Omega}$ for clarity. b.) Heatmap of the simulated signal.}
	\label{fig:sim_ee_iv}
\end{figure}

Under these assumptions, the low energy behavior is similar to the elastic case for bias up to $3\Delta_0$, where the pair potential now starts to drop approximately linearly, as seen in the solid line of Fig.~\ref{fig:sim_ee_delta}.a. Correspondingly, Fig.~\ref{fig:sim_ee_delta}.b shows this via the distribution function $f_\mathrm{L}$ as a function of energy for different bias conditions. Naturally, $f_\mathrm{L}$ is generally largest just above $\Delta$, and in both models decreases with increasing energy. Without inelastic scattering (dashed lines), the distribution function is nearly constant for large energy until it drops quickly to zero at $\epsilon=eV_\mathrm{inj}/\Delta_0$, reflecting the Fermi edge shifted by the applied bias. With inelastic scattering included however, $f_\mathrm{L}$ is enhanced at low energy due to thermalization but drops suddenly at energies about $\epsilon=3\Delta$ (indicated by arrows). This occurs as, for $\epsilon\geq3\Delta$, quasiparticles can release $\epsilon\geq 2\Delta$ by breaking Cooper pairs. As a consequence, the population of low-energy quasiparticles ($\epsilon\gtrsim\Delta$) increases, and the pair potential is further reduced. This is significant as low-energy quasiparticles are most effective in reducing $\Delta$.

By a similar method, Fig.~\ref{fig:sim_ee_iv} shows the asymmetric part $g_\mathrm{sa}$ of the non-local signal simulated using the self-consistent $\Delta$ shown in Fig.~\ref{fig:sim_ee_delta}(a) including inelastic scattering. The sharp peaks at $e|V_\mathrm{inj}|\approx\Delta_0$ correspond to the sharp initial drop of $\Delta$, while the additional signal above 
$3\Delta_0$ corresponds to the linear slope of $\Delta$ at high bias. Note that this simulation only takes into account the modulation of the local detector current due to the change $dN/dV_\mathrm{inj}$ of the density of states. Therefore, the signal is also odd in $V_\mathrm{det}$. In general, the behavior closely matches that observed in the non-local conductance (Fig.~\ref{fig:FullHeatmaps}), with the exception that the sign of the conductance does not reverse as seen in experiment for injector biases $\Delta<|eV_\mathrm{inj}|\leq3\Delta$ and $|eV_\mathrm{det}|>\Delta$. As such, the effect is not completely modeled under the assumptions made in our treatment of the kinetic equations, leaving open the possibility for future developments of this model.

\begin{figure}[tb!]
	\centering
    \ifdefined\withfigures
	\includegraphics[width=0.95\linewidth]{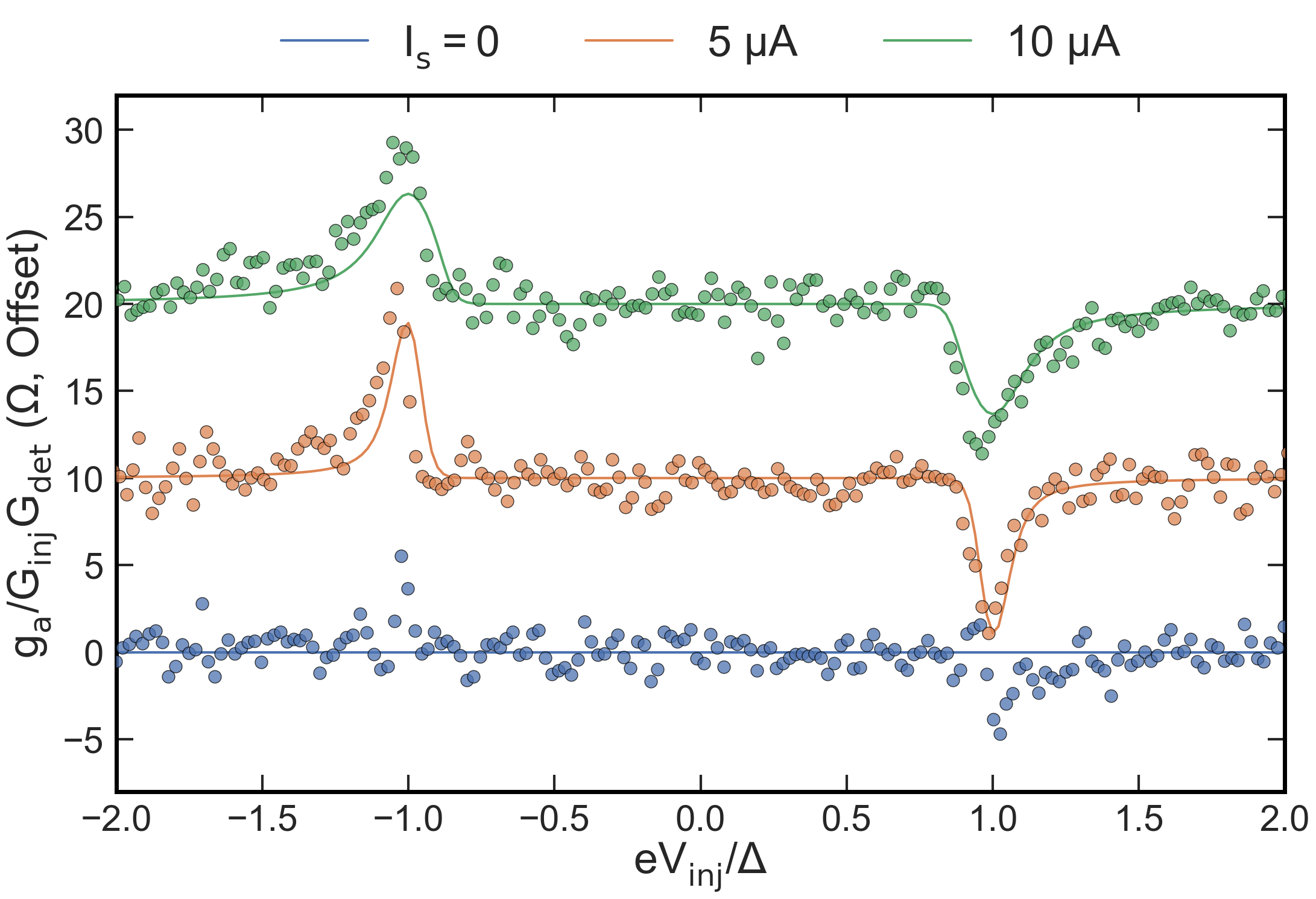}\\
    \fi
	\caption{Asymmetric part $g_\mathrm{a}$ of the non-local conductance as a function of injector bias $V_\mathrm{inj}$ for different supercurrents $I_\mathrm{s}$. Symbols are experimental data, lines are numerical simulations. The data are offset successively by $10~\mathrm{\Omega}$ for clarity.}
	\label{fig:sim_coupled_modes}
\end{figure}

As a final comparison to our model, the asymmetric part $g_\mathrm{a}$ (with respect to current only) of the non-local signal is shown in Fig.~\ref{fig:sim_coupled_modes} for three different supercurrent biases. Symbols are the experimental data, lines are numerical simulations including the coupling term, but without inelastic scattering. All parameters are taken from the fits shown in Fig.~\ref{fig:fit_fT_vs_Is}, with no free parameters left. Both the peak position and magnitude match closely with the expected kinetic mixing effect on the non-local conductance.

\section{\label{s:Conclusion}Conclusion}

The effect of inelastic scattering on non-local quasiparticle transport experiments was considered via comparison between experiment and quasiclassical simulations of the Usadel equations, both with and without inelastic electron-electron scattering. From this analysis, we are able to explain the predominantly anti-symmetric in bias contributions to the non-local conductance in this type of experiment by energy imbalance that is enhanced by the injection of higher energy quasiparticles with $\epsilon\geq3\Delta$. These quasiparticles lead to additional pairbreaking with a sharp energy onset that has not been previously observed in similar experiments. By fitting to our results, we ascertain estimates of the inelastic scattering rate for this process.

Via the design of our device, we could also consider the effect of supercurrent bias, which serves to kinetically couple the Usadel equations and leads to additional depairing. Experimentally this manifests as an evolution of the components of the non-local conductance of different symmetries with respect to bias polarity and supercurrent direction. Such effects are explained by the symmetries of the $f_\mathrm{L/T}$ modes, the depairing rate $\zeta$ and the form of the kinetic equations. Our results demonstrate a viable way of studying the kinetic mixing of these transport modes via symmetry decomposition of the non-local conductance which has mostly been considered only in spin-polarized transport devices. Several of the additional non-equilibrium effects observed, such as the roughly linear suppression of $I_c$ under bias and the non-BCS density of states seen during the application of a large supercurrent are also of interest, particularly for the research involving similarly biased superconducting nanowire devices.

As a final conclusion, we continue to observe additional anti-symmetric behavior inconsistent with the present model provided by the Usadel equations, under the assumptions that the non-local transport is only described in terms of the $f_\mathrm{L/T}$ modes and with equilibrium reservoirs. The effect is associated with anti-symmetric non-local conductance for approximately equal in magnitude detector and injector biases above the gap, which are of the opposite sign as those produced by the $f_\mathrm{L}$ mode contribution of our model. Based on our observations, it is likely that either the commonly used assumption of equilibrium behavior in the normal metal injectors should be reconsidered, or that there are more complex or coherent transport effects which occur during simultaneous current injection which are not considered by to our current model of quasiparticle tunneling.

\begin{acknowledgments}
This material is based upon work supported by the U.S. Department of Energy, Office of Science, National Quantum Information Science Research Centers, Superconducting Quantum Materials and Systems Center (SQMS) under the contract No. 89243024CSC0000020. This work was supported by the Helmholtz Association through the program NACIP. K.M.R. additionally acknowledges the support of the Colorado School of Mines and Prof. W. Van De Pontseele during the completion of this manuscript. 
\end{acknowledgments}

\appendix

\section{Non-local Measurements with an In-plane Magnetic Field}\label{App:MagField}

\begin{figure}[t!]
    \centering
    \ifdefined\withfigures
    \includegraphics[width=0.95\linewidth]{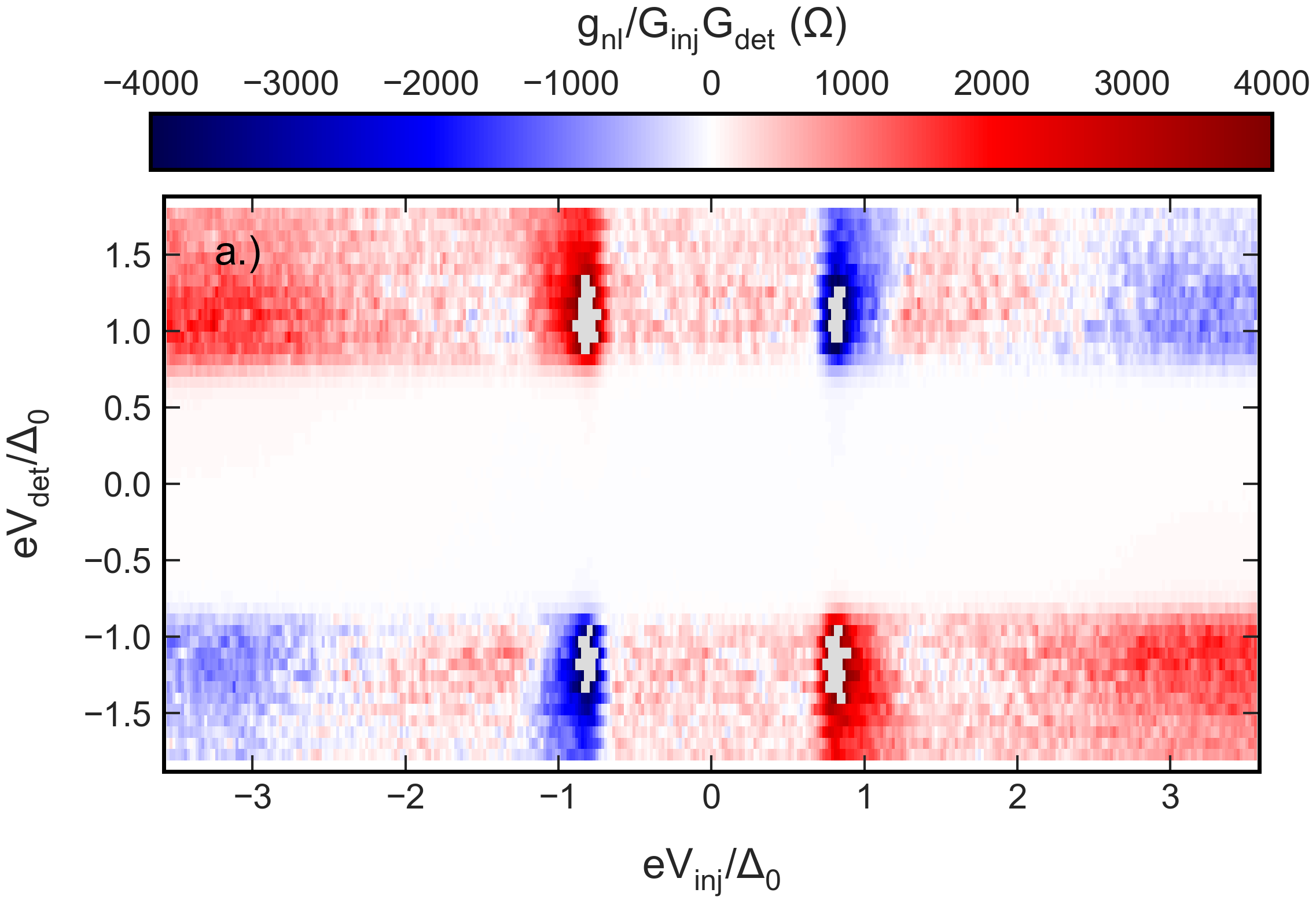}\\
    \includegraphics[width=0.95\linewidth]{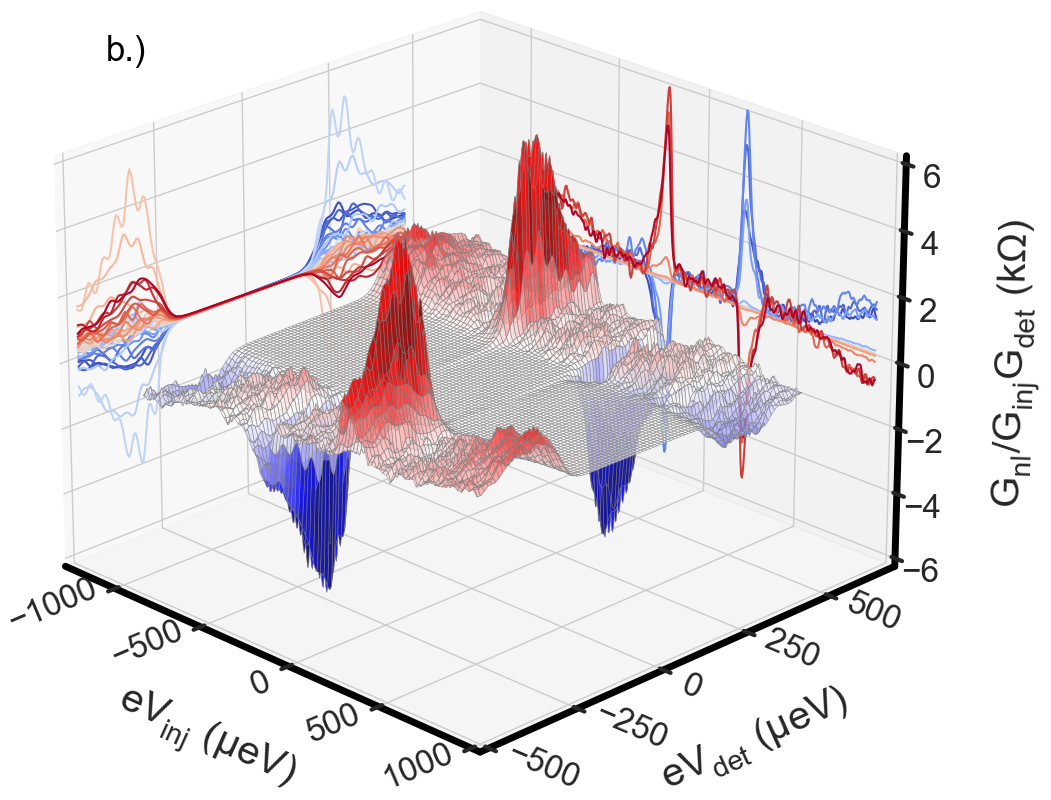}
    \fi
    \caption{Non-local conductance data taken under an applied \SI{1}{T} in plane magnetic field along the length of the wire for the $A/B$ junction pair. Bias voltages are normalized to the zero field equilibrium gap value. a.) Heatmap presentation, presented over the same limited range as in Fig.~\ref{fig:FullHeatmaps}. The grey regions correspond to data outside the color-map. b.) Same data, presented as a 3D contour over its full data range, along with projections indicating the general shape of the data taken during biasing of either $V_\mathrm{inj}$ or $V_\mathrm{det}$.  }
    \label{fig:FullHeatmaps_1Tesla}
\end{figure}

\begin{figure}[tb!]
    \centering
    \ifdefined\withfigures
    \includegraphics[width=0.96\linewidth]{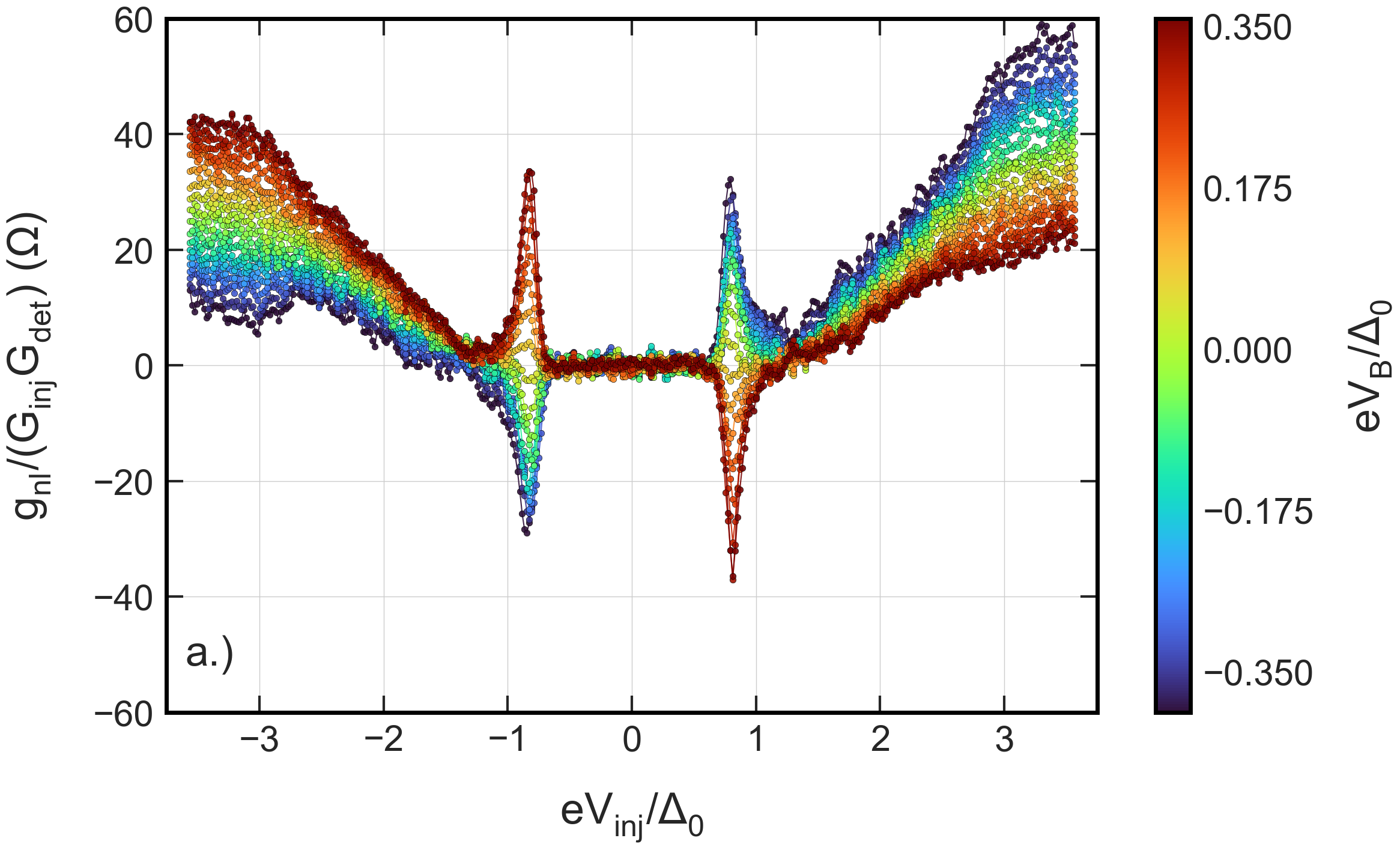}\hfill
    \vspace{1em}
    \includegraphics[width=0.96\linewidth]{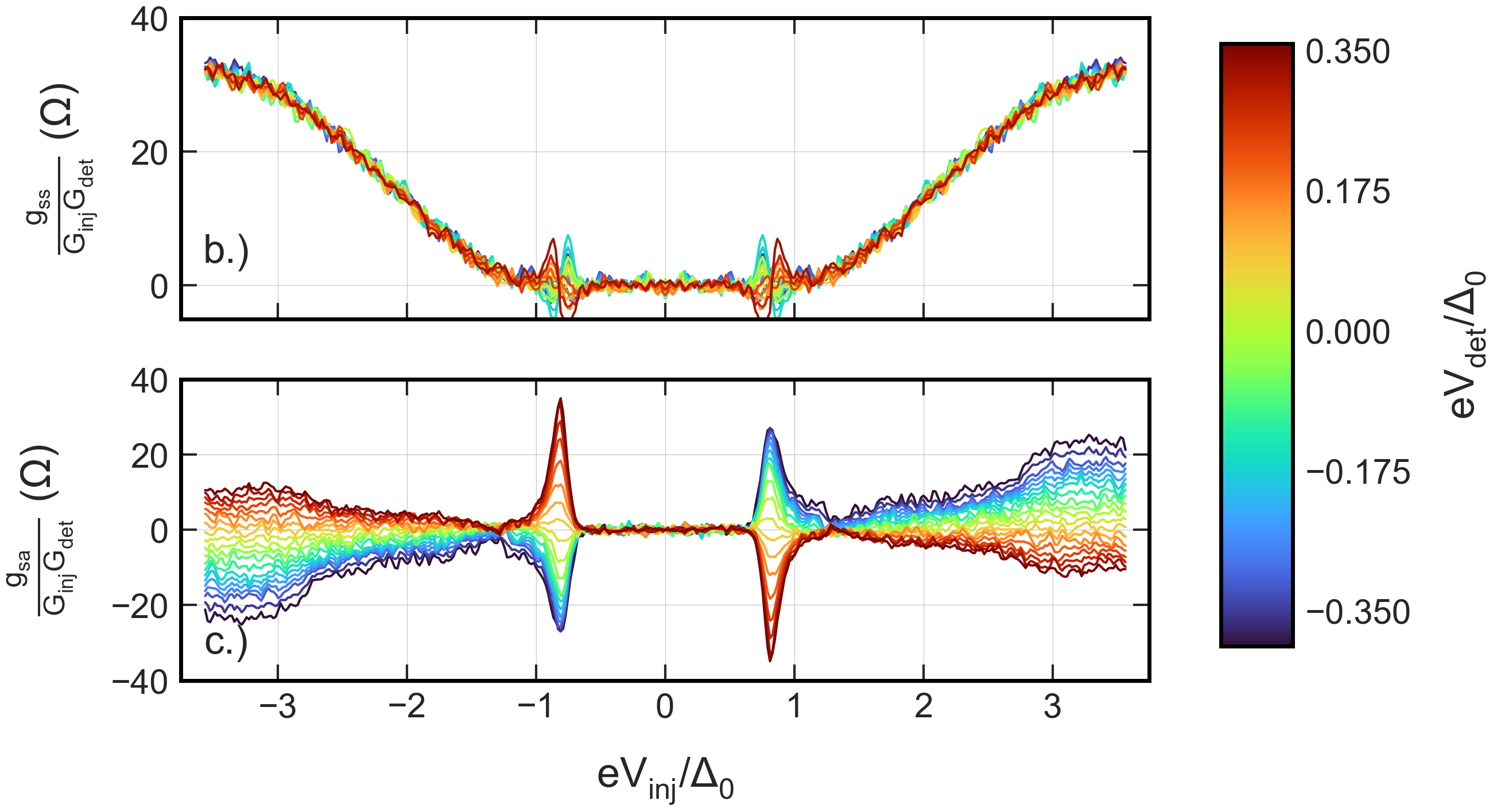}\hfill
    \fi
    \caption{Sub-gap region of the non-local conductance data taken under an applied \SI{1}{T} in plane magnetic field along the length of the wire. Bias voltages are normalized to the zero field equilibrium gap value. a.) Normalized non-local conductance between the $A/B$ junctions.  b.)+c.) Corresponding decompositions of the non-local dual-biased conductance data into its components symmetric/antisymmetric with respect to the injector voltage $V_\mathrm{inj}$. }
    \label{fig:SubGapBiasDecomp_1Tesla}
\end{figure}

Additional measurements were performed under an in plane magnetic field, oriented along the direction of the central superconducting wire using the $A/B$ junction pair. In thin films, the field penetration in this orientation is such that a Zeeman splitting of the spin-up and spin-down quasiparticle bands is expected to occur\cite{fulde_high_1973,beckmann2024}. In the device considered in the main text, we did not observe clear Zeeman splitting of the density of states at our maximum field of \SI{1}{T}, possibly due to an unexplained degree of spin-orbit scattering which smears out the density of states. Nevertheless, the non-local conductance is modified considerably, being both reduced in magnitude (likely by quasiparticle trapping by vortices) and ``stretched'' abbout the line $|V_\mathrm{inj}|=|V_\mathrm{det}|$ as shown in Fig.~\ref{fig:FullHeatmaps_1Tesla}. The effect in the subgap region (Fig.~\ref{fig:SubGapBiasDecomp_1Tesla}) is more akin a strong smearing of the charge and energy imbalance components, where now the inelastic scattering is broadened and no longer becomes sharply peaked at $\epsilon=3\Delta$. This could be due again to the contribution of trapping via vortices, or by the effect of Zeeman splitting of the energy levels involved during inelastic electron-electron scattering.

\section{Details of the Model and Fits}\label{App:Model}

For modeling the data, we have assumed an injector at the center of the wire ($x=0$), and equilibrium reservoirs at the end. We have used the following parameters (in addition to the ones given in Table~\ref{tab:Junctions}):

\paragraph*{Fig.~\ref{fig:fit_fT_vs_Is}---}
The kinetic equation reduces to
\begin{equation}
\mathcal{D}_\mathrm{T}\partial_x^2f_\mathrm{T} = \mathcal{R}_\mathrm{T} f_\mathrm{T}
\end{equation} with $\mathcal{D}_\mathrm{T}=\mathrm{Re}(\greensg)^2+\mathrm{Re}(\greensf)^2$ and $\mathcal{R}_\mathrm{T}=2\mathrm{Re}(\greensf)\Delta$. The solution for the spectral non-local conductance is
\begin{equation}
	g_\mathrm{nl}(\epsilon)=\frac{\kappa_\mathrm{I}N}{\kappa_\mathrm{I}N+\mathcal{D}_\mathrm{T}/\lambda_\mathrm{T}}N\exp(-|x|/\lambda_\mathrm{T})
\end{equation}
with $\lambda_\mathrm{T}=\sqrt{\mathcal{D}_\mathrm{T}/\mathcal{R}_\mathrm{T}}$. The convolution of $g_\mathrm{nl}(\epsilon)$ with the derivative of the Fermi function is the actual non-local conductance. To account for the left and right branches of the wire in parallel, we set $\kappa_\mathrm{I}=G_\mathrm{inj}R_\xi/2$.

\paragraph*{Figures~\ref{fig:sim_ee_delta} and \ref{fig:sim_ee_iv}---} 
The kinetic equation reduces to
\begin{equation}
	\mathcal{D}_\mathrm{L}\partial_x^2f_\mathrm{L} = \mathcal{I}_\mathrm{L}
\end{equation}
with $\mathcal{D}_\mathrm{L}=\mathrm{Re}(\greensg)^2-\mathrm{Im}(\greensf)^2$. The collision integral $\mathcal{I}_\mathrm{L}$ is given by Eq.~(16) of \cite{bobkova2016}. We solve the equation on an equidistant energy grid with spacing $\delta\epsilon=1/40$ using a standard numerical routine \cite{shampine2006}. The collision integral couples all energies, i.e., the problem has to be solved on the whole grid simultaneously. To keep the problem tractable, we choose a relatively small $\omega_\mathrm{D}=10$ and set $\lambda=1/\mathrm{arcsinh}(\omega_\mathrm{D})$ to ensure $\Delta=1$ at $t=0$. We had to choose a relatively large $\Gamma=0.005$ to smooth out the gap features to obtain reasonable convergence. Also, we only solve for one half of the wire with double cross section, i.e., $\kappa_\mathrm{I}=G_\mathrm{inj}R_\xi/2$. We still assume $\Delta(x)\equiv \mathrm{const.}$, and insert the distribution function at $x=0$ into the self-consistency equation Eq.~(\ref{eqn:sc}). We iterate until self-consistence is obtained for each bias point on a grid from $0\leq eV_\mathrm{inj}\leq 5$ with spacing $0.1$. We then calculate $I_\mathrm{det}(V_\mathrm{inj},V_\mathrm{det})$ and obtain $dI_\mathrm{det}/dV_\mathrm{inj}$ numerically. We neglect the dependence of $\Delta$ on $V_\mathrm{det}$, since it should not affect the derivative w.r.t. $V_\mathrm{inj}$ much.

\paragraph*{Fig.~\ref{fig:sim_coupled_modes}---}
We solve the coupled kinetic equations Eqs.~\ref{eqn:kinfL} and \ref{eqn:kinfT} simultaneously without inelastic collisions \cite{maier2023}. Here, we have to solve the problem for both branches of the wire, since the supercurrent along the wire flows past the injector junction. Without inelastic scattering, we can solve for each energy separately. We choose $\delta\epsilon=1/50$ up to $|\epsilon|=3\Delta_0$.

\bibliography{references}% Produces the bibliography via BibTeX.

\end{document}